\documentclass{emulateapj}
\slugcomment{{\sc Accepted for publication in ApJ.}} 

\usepackage{amsmath}

\newcommand{\pasa}{PASA}
\newcommand{\nar}{NAR}

\shortauthors{F\"orster, Gonz\'alez--Gait\'an, Folatelli \& Morrell}
\begin{document}

\title{On the Lira law and the nature of extinction towards Type Ia
  Supernovae.}  \author{Francisco F\"orster$^{1,2}$, Santiago
  Gonz\'alez--Gait\'an$^1$, Gast\'on Folatelli$^3$, Nidia Morrell$^4$}
\affil{$^1$ Departamento de Astronom\'\i a, Universidad de Chile,
  Casilla 36-D, Santiago, Chile} \affil{$^2$ Center for Mathematical
  Modelling, Universidad de Chile, Avenida Blanco Encalada 2120 Piso
  7, Santiago, Chile} \affil{$^3$Kavli Institute for the Physics and
  Mathematics of the Universe, Todai Institutes for Advanced Study
  (TODIAS), University of Tokyo, 5-1-5 Kashiwanoha, Kashiwa, Chiba
  277-8583, Japan} \affil{$^4$Las Campanas Observatory, Carnegie
  Observatories, Casilla 601, La Serena, Chile}

\begin{abstract}

We have studied the relation between the color evolution of Type Ia
Supernovae (SNe Ia) from maximum light to the \emph{Lira law} regime
and the presence of narrow absorption features. Based on a nearby
sample of 89 SNe Ia we have found that the rate of change of $B-V$
colors at late phases (between 35 and 80 days after maximum) varies
significantly among different SNe Ia.  At maximum light, faster Lira
law $B-V$ decliners have significantly higher equivalent widths of
blended Na I D1 \& D2 narrow absorption lines, redder colors and lower
$R_{V}$ reddening laws. We do not find faster Lira law $B-V$ decliners
to have a strong preference for younger galaxy environments, where
higher interstellar material (ISM) column densities would be
expected. We interpret these results as evidence for the presence of
circumstellar material (CSM). The differences in colors and reddening
laws found at maximum light are also present 55 days afterward, but
unlike the colors at maximum, they show a significant variation among
different host galaxy morphological types. This suggests that the
effect of ISM on the colors is more apparent at late times. Finally,
we discuss how the transversal expansion of the ejecta in an
inhomogeneous ISM could mimic some of these findings.

\end{abstract}

\keywords{supernovae: general --- distance scale}

\section{INTRODUCTION}

Type Ia supernovae (SNe Ia) are thought to be the thermonuclear
explosion of white dwarf stars. They are one of the main sources of
iron in the Universe \citep[e.g.][]{1986A&A...154..279M} and are key
tools for cosmological distance determinations
\citep{1993ApJ...413L.105P, 1996AJ....112.2408H}, leading to the
discovery of the acceleration of the Universe \citep{riess98,
  perl99}. Despite much observational and theoretical effort, we lack
a clear understanding about their progenitors \citep[for reviews,
  see][]{2000ARA&A..38..191H, 2012NewAR..56..122W}.

Most favored progenitor scenarios involve a carbon--oxygen white
dwarf (CO WD) in a binary system. The CO WD could either ignite at the
Chandrasekhar mass after accreting matter stably from a companion in a
delayed detonation or pure deflagration explosion \citep[Chandrasekhar
  mass -- single degenerate scenario, $M_{Ch}$ --
  SD:][]{1996ApJ...470L..97H, 1999ApJ...519..314H,
  1999ApJ...522..487H, 1997A&A...322L...9L, 2000A&A...362.1046L,
  2004MNRAS.350.1301H, 2009MNRAS.395.2103M, 2009ApJ...701.1540W,
  1991A&A...245L..25K, 2007A&A...464..683R, 2012arXiv1210.5243K}, at a
mass below the Chandrasekhar mass if accretion is unstable and a
surface detonation is strong enough to trigger a central detonation
\citep[sub--Chandrasekhar mass -- single degenerate scenario, $M_{Ch}$
  -- SD:][]{1982ApJ...257..780N, 1986ApJ...301..601W,
  2010ApJ...714L..52S, 2010ApJ...719.1067K}, or after the merger with
a degenerate companion in a double degenerate scenario if the
detonation can occur in a dynamical time--scale \citep[violent double
  degenerate merger, violent DD:][]{1984ApJS...54..335I,
  1984ApJ...277..355W, 2012ApJ...747L..10P}. Some observational tests
for these scenarios can be: 1) the presence or absence of low velocity
circumstellar material left from the accretion process or from
previous nova explosions, 2) the presence or absence of a companion
star, 3) the SN rate delay--time distribution, which is generally
found to be inversely proportional to the time since formation
\citep[for a review, see][]{2012PASA...29..447M} and is generally
believed to favor DD scenarios \citep[although
  see][]{2009ApJ...707.1466K}, and 4) the degree of neutronization in
the central regions due to different central densities at ignition
\citep[e.g.][]{2007ApJ...661..995G}.

Although the lack of evidence for shocked companion stars of the
progenitor CO WDs, shocked presupernova winds or companion
pre--explosion images is more generally consistent with DD scenarios
\citep{2011Natur.480..348L, 2011Natur.480..344N, 2012ApJ...744L..17B,
  2012ApJ...753...22B, 2012Natur.481..164S, 2012ApJ...747L..19E},
there are different results that suggest the presence of circumstellar
material (CSM) around SN Ia progenitors, which is generally expected
in SD scenarios \citep[or slow DD mergers,
  see][]{2010ApJ...725..296F}. These consist of optical colors that
favor scattering in nearby material \citep{2005ApJ...635L..33W,
  2008ApJ...686L.103G, 2011ApJ...735...20A}, narrow absorption lines
that show variability \citep{2007Sci...317..924P, 2009ApJ...693..207B,
  2009ApJ...702.1157S}, are blueshifted \citep{2011Sci...333..856S,
  2012ApJ...752..101F}, or that correlate with certain SN ejecta
viewing angles \citep{2012ApJ...754L..21F}. This evidence is
contrasted by the absence of X--ray or radio emission from shocked CSM
in the nearby SN 2011fe \citep{2012ApJ...746...21H}, but is supported
by recent observations of SN Ia PTF11kx \citep{2012Sci...337..942D},
which was probably surrounded by narrow nova shells ejected before
explosion, and the sample of SNe Ia with CSM interaction recently
identified by \citet{2013arXiv1304.0763S}.

If some SNe Ia are surrounded by significant amounts of CSM, it would
be expected that the light from those SNe would be scattered away of
the line of sight by intervening dust, causing extinction and
reddening, but also that it would be scattered into the line of sight
by nearby dust, i.e. multiple scattering processes or light echoes
\citep[e.g.][]{1986ApJ...308..225C, 2005ApJ...635L..33W,
  2008ApJ...686L.103G}. The typical delay of these light echoes would
depend on the typical distance to the closest surviving dust after
explosion, which for typical distances of about $0.005-0.1$ pc
\citep[see e.g.][]{2007Sci...317..924P, 2009ApJ...702.1157S} would
correspond to approximately 6 to 112 days of light
travel--time. Assuming that the light echoes and/or dust sublimation
at later times change the SN light curves at these time--scales after
maximum\footnote{We define maximum as the maximum of the SN light
  curve in $B$--band.}, we have done a systematic study of the
evolution of late time colors of SN Ia during the Lira law regime
\citep{1996MsT..........3L, 1999AJ....118.1766P}, between 35 and 80
days after maximum. Given that SN colors evolve in an approximately
homogeneous fashion during this phase, we try to detect the presence
of multiple scattering processes, or light echoes, studying the
evolution of colors in a big sample of nearby SNe Ia and its relation
with the presence of narrow absorption features.

In order to distinguish between interstellar material (ISM) and
circumstellar material (CSM) we have studied the relation between the
equivalent widths of blended Na I D1 \& D2 narrow absorption features,
obtained from mid--resolution spectra using the method described in
\citealt[][hereafter FG12]{2012ApJ...754L..21F}; with the Lira law
$B-V$ decline rates. Extinction by ISM should in principle only shift
the $B-V$ colors to higher values, without significant changes of the
Lira law $B-V$ decline rate, whereas CSM can both shift the $B-V$
colors to higher values and change the Lira law $B-V$ decline rate,
making it generally faster as the distance to the CSM is reduced
\citep{2011ApJ...735...20A} or if CSM dust is destroyed at late
times. Given that systems with significant CSM should have a different
extinction law \citep{2005ApJ...635L..33W, 2008ApJ...686L.103G}, we
also study the relation between $B-V$ and $V-i$ colors at maximum and
55 days after maximum for different Lira law $B-V$ decline rates in
our sample. To discard that SN light curves have different intrinsic
decline rates for different galaxy populations we have also performed
several environmental tests studying the distributions of host galaxy
morphological type, host galaxy inclination, stretch or $\Delta
m_{15}$ of different Lira law $B-V$ decline rates.

Energy deposition during the Lira law regime is dominated by the
$\beta$--decay of $^{56}$Co into $^{56}$Fe, via electron captures
(81\%) or positron emissions (19\%). At 35 days after maximum, energy
deposition is dominated by the scattering of $\gamma$--rays produced
in the primary channel. At 80 days after maximum, the ejecta becomes
transparent to $\gamma$--rays and the kinetic energy from positrons
produced in the secondary channel dominates the energy deposition
\citep{1999ApJS..124..503M}. This means that a transition from
$\gamma$--ray to positron energy deposition occurs during the Lira law
regime, which may be linked to different bolometric decline
rates. However, to our knowledge there is not a clear connection
between the bolometric decline rate with the $B-V$ decline rates
observed during the Lira law.

Finally, given the connection found by FG12 between narrow absorption
features and the nebular velocity shifts introduced by
\citet{2010ApJ...708.1703M}, interpreted as evidence for CSM in
positive nebular velocity shift SNe Ia and a general asymmetry of the
CSM post--explosion, we have tested whether there is any correlation
between nebular velocity shifts and the observed Lira law $B-V$
decline rates.

\section{SAMPLE SELECTION AND ANALYSIS} \label{sec:sample}

We have selected those SNe Ia that have spectra in the Center for
Astrophysics (CfA) public sample of SNe Ia \citep{2012AJ....143..126B}
or in the Carnegie Supernova Program (CSP) sample of nearby SNe Ia
\citep{F+2013}. For every spectrum of every SN we have run an
automatic Python script to measure equivalent widths and associated
errors of blended Na I D1 \& D2 narrow absorption features as
described in FG12. These values were combined using a weighted average
to obtain one equivalent width and associated error per SN, assigning
an equivalent width of zero to those SNe with negative
measurements. We then selected those SNe with final equivalent width
errors below 0.6 \AA.

For the photometry, we used data from either CfA3 and CfA4
\citep{2009ApJ...700..331H, 2012ApJS..200...12H}, CSP
\citep{2010AJ....139..519C, 2010AJ....140.2036S, 2011AJ....142..156S}
and other sources (see Table~\ref{tab:other}). When multiple sources
were available, we use either only CSP or only CfA data depending on
the numbers of photometric points available. We corrected for Milky
Way extinction using the maps of \citet{1998ApJ...500..525S} and
performed K--corrections and S--corrections to the natural rest--frame
system of the Las Campanas Observatory (LCO)
\citep{2011AJ....142..156S} using photometry--adjusted spectral
template series from \citet{2007ApJ...663.1187H}. In those cases where
the natural system of the instrument is not provided and the
photometry is only given in the system of \cite{1992AJ....104..372L},
we transform from there to the LCO system. We selected only those SNe
Ia with good photometric coverage during the Lira law regime (between
35 and 80 days after maximum), which we defined as those SNe having at
least three simultaneous $B$ and $V$ photometric data points with a
minimum separation of 25 days between the first and last observations.

\begin{table*}[h!t]
 \centering
 \caption{Nearby SNe Ia used in this analysis besides data from the
   Center for Astrophysics \citep[CfA3 and
     CfA4:][]{2009ApJ...700..331H, 2012ApJS..200...12H} and the
   Carnegie Supernova program \citep[CSP:][]{2010AJ....139..519C,
     2010AJ....140.2036S}. Note that we do not include data from the
   YALO telescope because of possible inconsistencies with its
   photometric system \citep{2003AJ....125..166K,
     2006AJ....131.2615P}.}
    \vspace{.5cm}
  \begin{tabular}{lc}
    \hline
    \hline
    \vspace{-.3cm} \\
    Name & Source \\
    \vspace{-.4cm} \\
    \hline
          SN 1993ae &  \citet{1999AJ....117..707R} \\
          SN 1994D &   \citet{1996MNRAS.281..263M, 2004MNRAS.349.1344A} \\
          SN 1994ae &  \citet{2005ApJ...627..579R, 2004MNRAS.349.1344A} \\
          SN 1995D  &  \citet{1999AJ....117..707R, 1996PASJ...48...51S} \\
          SN 1995al &  \citet{1999AJ....117..707R} \\
          SN 1995bd &  \citet{1999AJ....117..707R, 2004MNRAS.349.1344A} \\
          SN 1996C  &  \citet{1999AJ....117..707R} \\
          SN 1997E  &  \citet{2006AJ....131..527J} \\
          SN 1997cw &  \citet{2006AJ....131..527J} \\
          SN 1998ab &  \citet{2006AJ....131..527J} \\
          SN 1998aq &  \citet{2005ApJ...627..579R} \\
          SN 1998bp &  \citet{2006AJ....131..527J} \\
          SN 1998dm &  \citet{2006AJ....131..527J, 2010ApJS..190..418G} \\
          SN 1998dh &  \citet{2006AJ....131..527J, 2010ApJS..190..418G} \\
          SN 1998ef &  \citet{2006AJ....131..527J, 2010ApJS..190..418G} \\
          SN 1998es &  \citet{2006AJ....131..527J} \\
          SN 1999aa &  \citet{2000ApJ...539..658K, 2004MNRAS.349.1344A, 2006AJ....131..527J} \\
          SN 1999ac &  \citet{2006AJ....131..527J, 2006AJ....131.2615P} \\
          SN 1999gd &  \citet{2006AJ....131..527J} \\
          SN 1999gh &  \citet{2006AJ....131..527J, 2010ApJS..190..418G} \\
          SN 1999dq &  \citet{2006AJ....131..527J} \\
          SN 1999gp &  \citet{2006AJ....131..527J, 2010ApJS..190..418G, 2001AJ....122.1616K} \\
          SN 2000B  &  \citet{2006AJ....131..527J} \\
          SN 2000cx &  \citet{2003PASP..115..277C, 2004MNRAS.349.1344A, 2004AA...428..555S, 2006AJ....131..527J} \\
          SN 2000dk &  \citet{2006AJ....131..527J, 2010ApJS..190..418G} \\
          SN 2000fa &  \citet{2006AJ....131..527J, 2010ApJS..190..418G} \\
          SN 2002aw &  \citet{2010ApJS..190..418G} \\
          SN 2002bo &  \citet{2004AJ....128.3034K, 2004MNRAS.348..261B, 2010ApJS..190..418G} \\
          SN 2002cs &  \citet{2010ApJS..190..418G} \\
          SN 2002dl &  \citet{2010ApJS..190..418G} \\
          SN 2003cg &  \citet{2006MNRAS.369.1880E, 2010ApJS..190..418G} \\
          SN 2003du &  \citet{2005AA...429..667A, 2007AA...469..645S, 2005ApJ...632..450L} \\
          SN 2004bg &  \citet{2010ApJS..190..418G} \\
          SN 2004bk &  \citet{2010ApJS..190..418G} \\
          \hline
  \end{tabular}
  \label{tab:other}
\end{table*}

For the final sample, we fitted a linear relation in the $B-V$ vs time
plane between 35 and 80 days after maximum. This led to two values and
associated errors for every SNe Ia: the $B-V$ color at 55 days and the
Lira law $B-V$ decline rate. We retained only those SNe Ia that had an
error associated with the Lira law $B-V$ decline rate smaller than 0.006
mag day$^{-1}$. 

For every SN in the sample we also fitted stretch and colors at
maximum using SiFTO \citep{2008ApJ...681..482C}, requiring at least
one data point between -15 and 7.5 days after $B$--band maximum and
another one between 2.5 and 35 days after $B$--band maximum in each
band, as well as $\Delta m_{15}$ values and $E(B-V)$ extinction values
using SNooPY \citep{2011AJ....141...19B}. For most purposes, stretch
and $\Delta m_{15}$ are equivalent quantities describing the light
curve shape diversity of SNe Ia, but we use both to account for
systematic effects in our analysis. We also obtained inclinations and
host galaxy morphological types using the Asiago catalogue
\citep{1999A&AS..139..531B} when available. In order to discard the
effects introduced by possibly very different populations in the SN Ia
sample, we have also removed sub--luminous SNe Ia, which we define as
those SNe Ia whose stretch is smaller than 0.7 and whose $B-V$ color
at maximum is higher than 0.4.

\begin{figure*}[h!]
  \centering 
  \hbox{
    \includegraphics[width=8cm,angle=0]{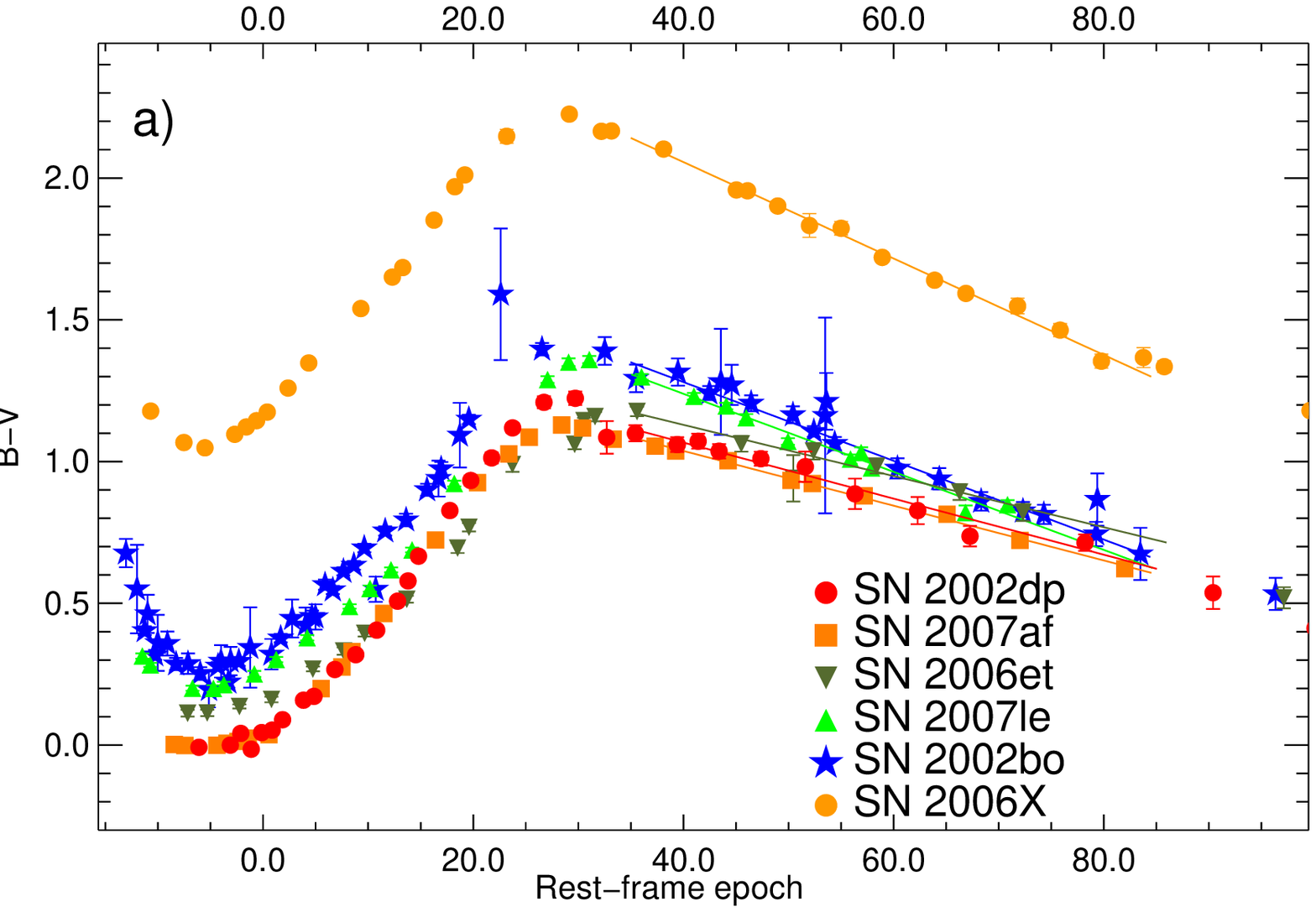}
    \includegraphics[width=8cm,angle=0]{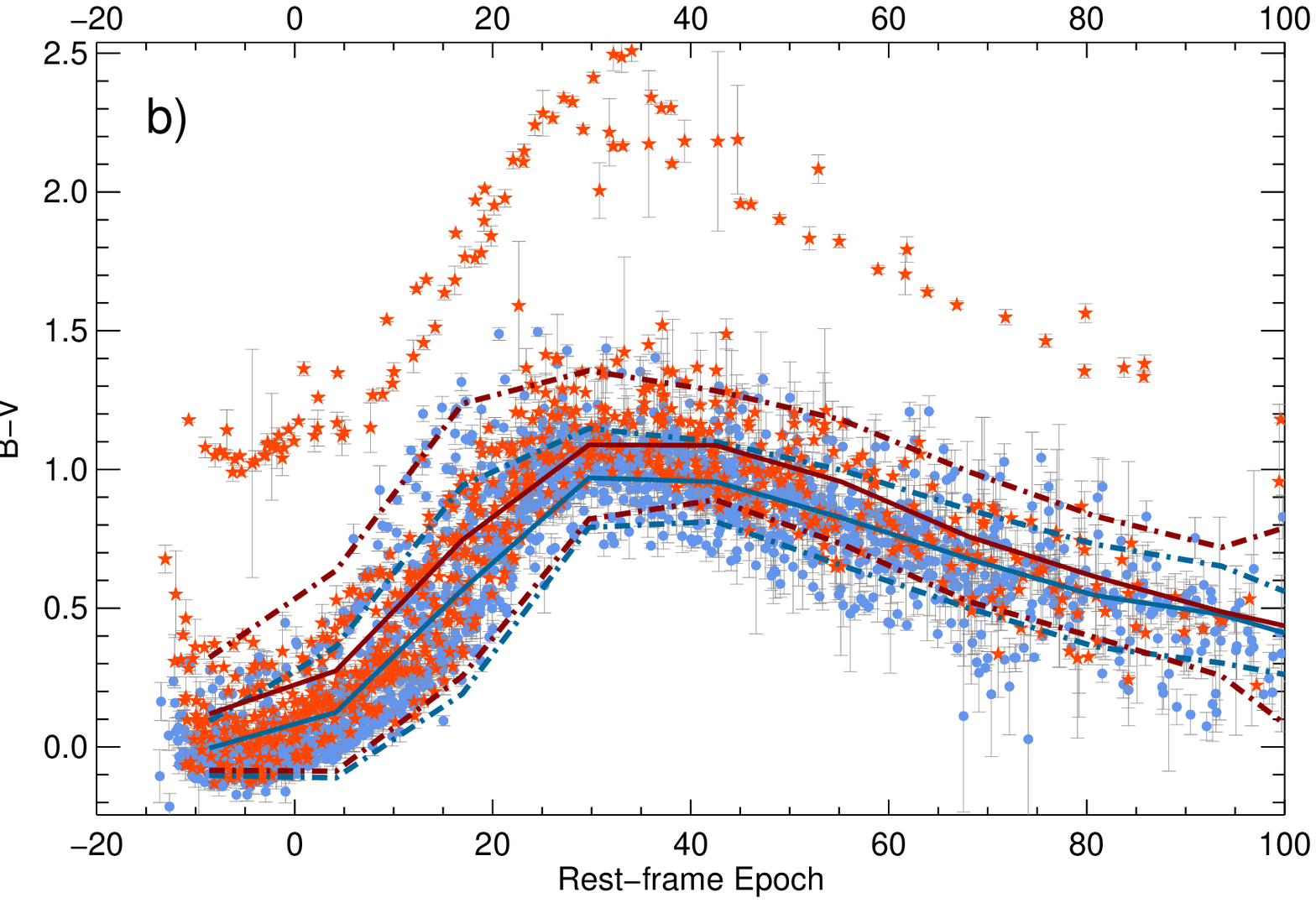}
  }
  \caption{a) $B-V$ evolution for selected SNe and b) for the entire
    sample defined in Section~\ref{sec:sample}, dividing the sample
    between SNe with Lira law $B-V$ decline rates faster (red stars)
    and slower (blue dots) than -0.013 mag day$^{-1}$. We plot the
    median evolution and observed dispersion for each sample with red
    and blue lines, respectively. It can be seen that faster Lira law
    $B-V$ decliners are redder at maximum and 55 days afterward.}
  \label{fig:BVsample}
\end{figure*}

In Figure~\ref{fig:BVsample} a) the $B-V$ color evolution for selected
SNe is shown and in b) for the entire sample used in this analysis,
separated by Lira law $B-V$ decline rate. Figure~\ref{fig:BVsample} a)
hints that different Lira law $B-V$ decline rates occur among
different SNe Ia. In the following we explore whether the Lira law
$B-V$ decline rate is related to other observational characteristics
in the sample.

\section{RESULTS}

\subsection{Lira law $B-V$ decline rates vs equivalent widths and colors}

\begin{figure*}[h!]
  \centering 
  \hbox{
    \includegraphics[width=8cm,angle=0]{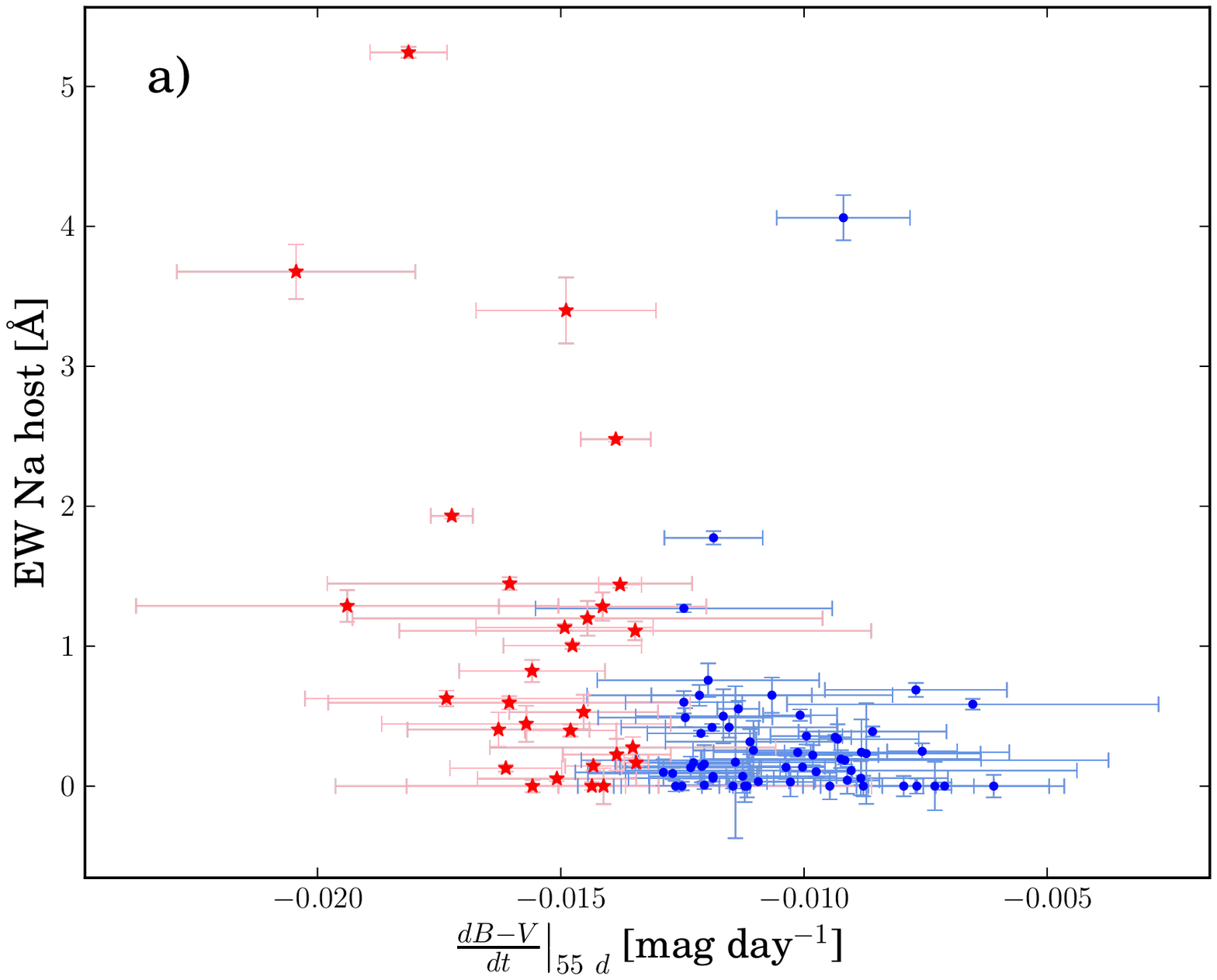}
    \includegraphics[width=8cm,angle=0]{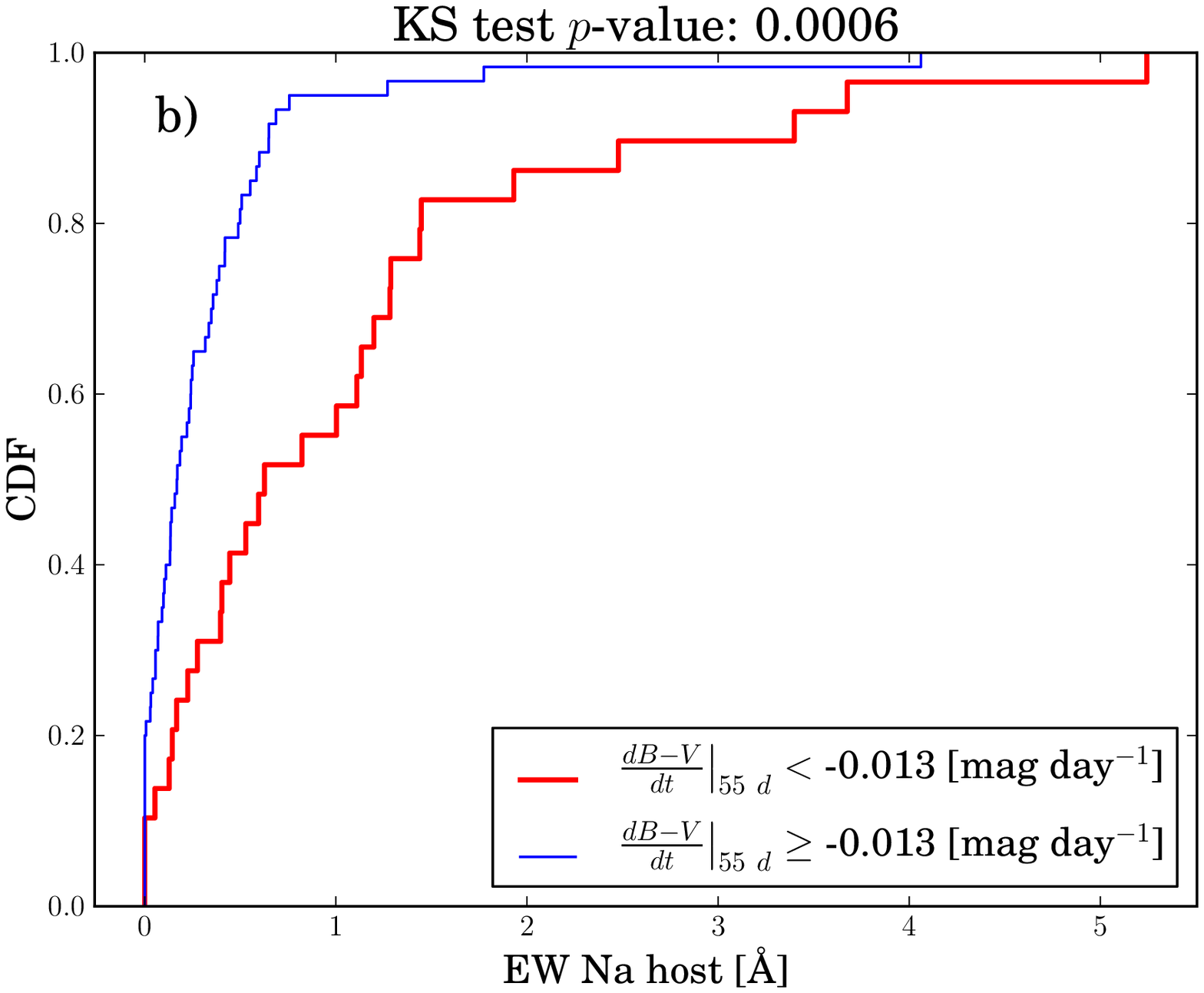}
  }
  \caption{a) Equivalent widths of narrow Na I D1 \& D2 blended
    absorption features vs the Lira law $B-V$ decline rate for fast
    (red stars) and slow (blue dots) Lira law $B-V$ decliners and b)
    the corresponding Kolmogorov Smirnov (KS) test for the cumulative
    distribution of equivalent widths of narrow Na I D1 \& D2 blended
    absorption features for fast (red) and slow (blue) Lira law $B-V$
    decline rates for the sample defined in
    Section~\ref{sec:sample}. The division between fast and slow Lira
    law $B-V$ decliners is defined to be at -0.013 mag day$^{-1}$
    (see Section~\ref{sec:stability}).}
  \label{fig:bBVew}
\end{figure*}

In Figure~\ref{fig:bBVew} we show the equivalent widths of blended
narrow Na I D1 \& D2 absorption features vs the Lira law $B-V$ decline
rate according to the sample defined in Section~\ref{sec:sample}. From
the Figure it is clear that the distribution of equivalent widths of
those SNe with faster Lira law $B-V$ decline rates is different from
that of SNe Ia with slower $B-V$ decline rates. If we divide the
sample by a Lira law $B-V$ decline rate of -0.013 mag day$^{-1}$, we
obtain with a Kolmogorov-Smirnov (KS) test that there is a probability
of 0.0006 that the resulting distributions of equivalent widths arise
from the same parent population. This division of the sample is based
on finding the Lira law $B-V$ decline rate that maximizes the
differences between fast and slow Lira law $B-V$ decliners (explained
in more detail in Section~\ref{sec:stability}) and it implies that
33\% of the SNe in the sample are defined as fast Lira law $B-V$
decliners.

\begin{figure*}[h!]
  \centering
  \hbox{
    \includegraphics[width=8cm,angle=0]{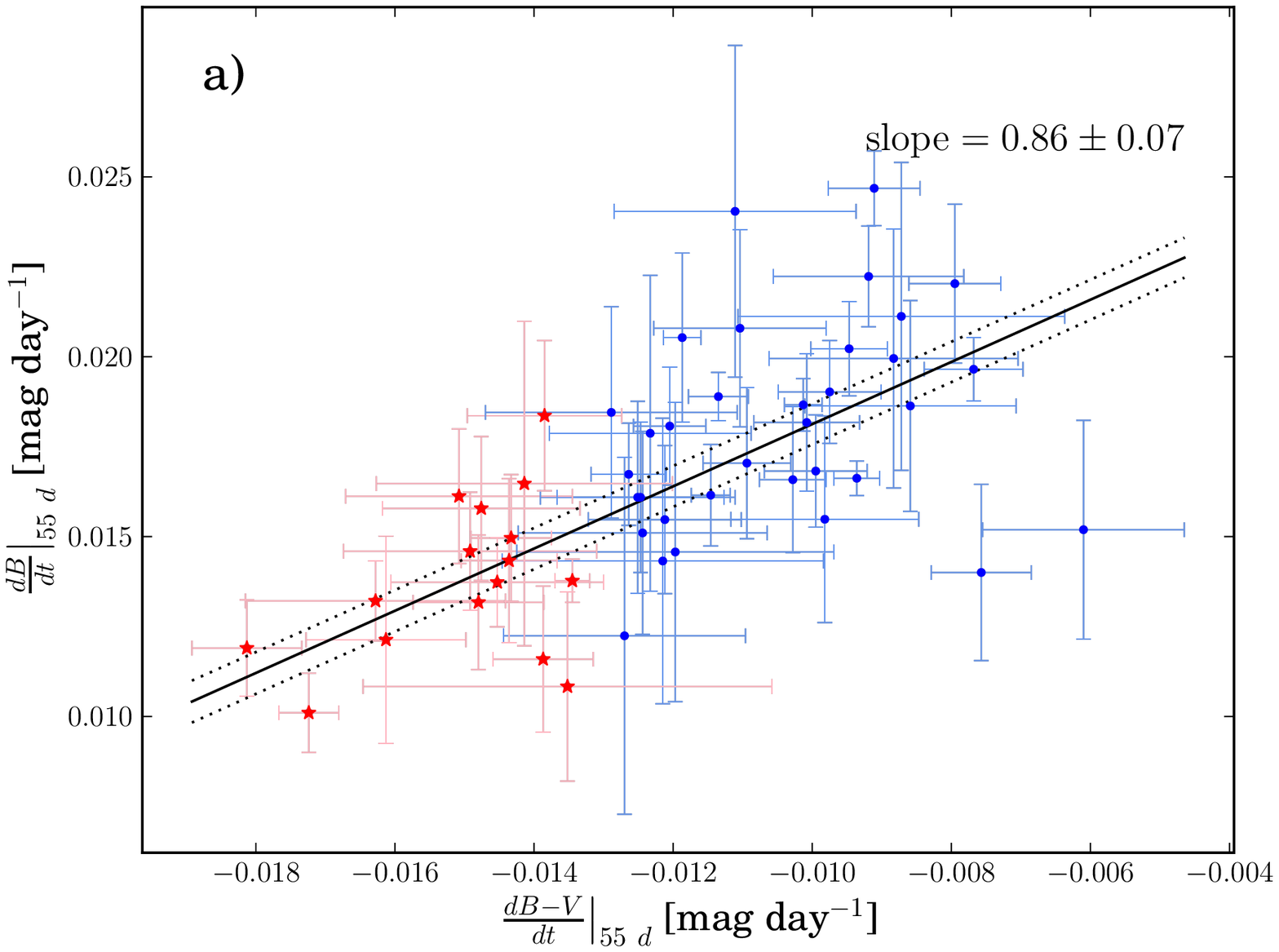}
    \includegraphics[width=8cm,angle=0]{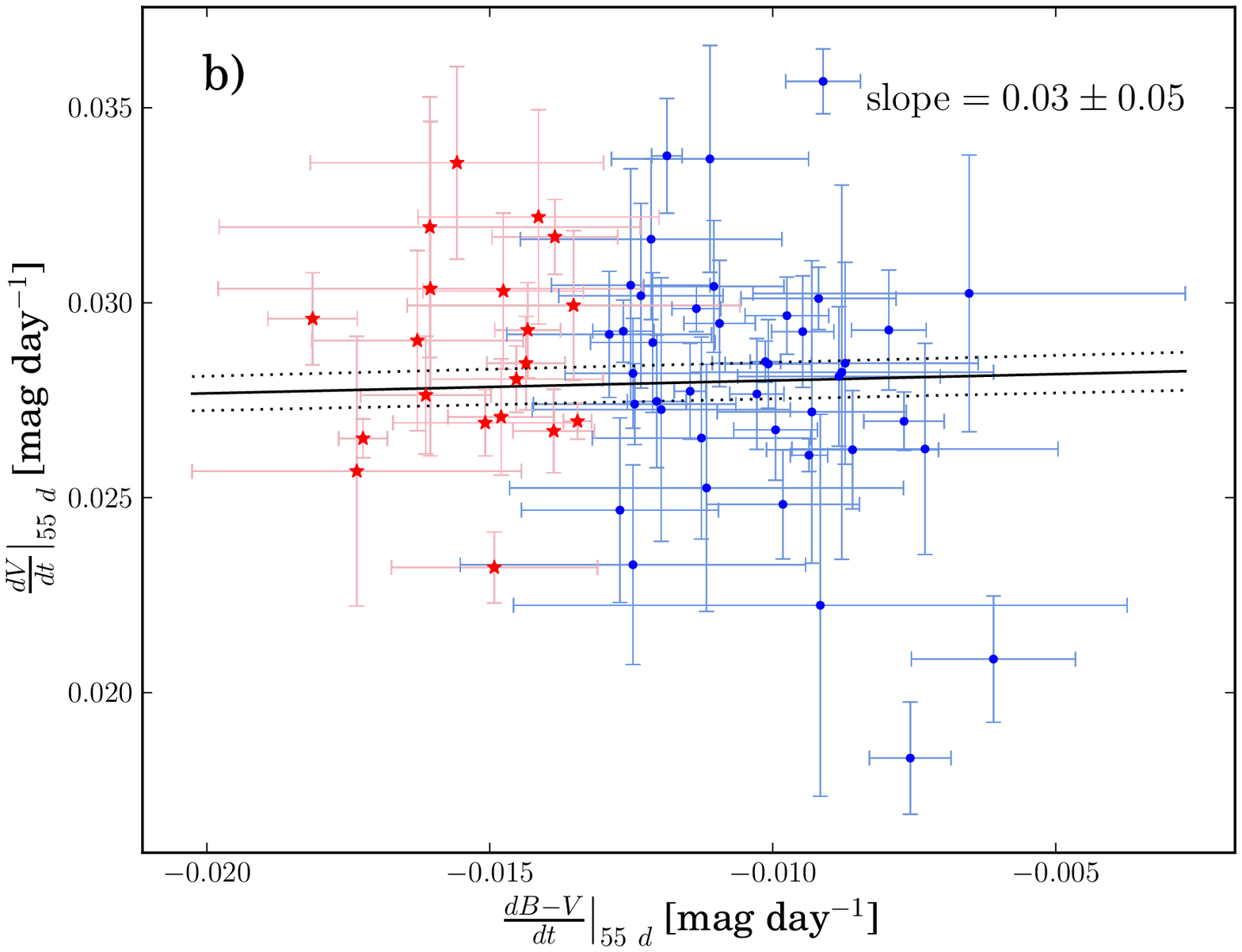} 
  }
  \caption{a) Lira law $B$--band decline rates vs Lira law $B-V$
    decline rates for fast (red stars) and slow (blue dots) Lira law
    $B-V$ decliners and b) Lira law $V$--band decline rates vs Lira
    law $B-V$ decline rates for fast (red stars) and slow (blue dots)
    Lira $B-V$ law decliners. We found that the diversity of $B-V$
    decline rates is mostly due to different $B$--band decline rates
    during the Lira law, and not to $V$--band decline rates
    variations, suggesting that the origin of the diversity has a
    larger effect at shorter wavelengths.}
  \label{fig:declines}
\end{figure*}

In order to study the effect of positron absorption, we investigate
whether the differences between fast and slow $B-V$ Lira law decliners
are related to their bolometric decline rates during the Lira law,
which we approximate by their $V$--band decline rates \citep[see
  e.g.][]{1997A&A...328..203C}.  We found that fast and slow Lira law
$B-V$ decliners do not have significantly different distribution of
$V$--band decline rates, with a KS test probability of 0.86 that their
distributions of $V$--band decline rates arise from the same parent
population. However, we find a significant difference between their
distributions of $B$--band decline rates, with a KS test probability
of only 0.0001 that their distributions of $B$--band decline rates
arise from the same parent population. Thus, the Lira law $B-V$
decline rate diversity is driven primarily by different $B$--band
decline rates, in the direction that fast Lira law $B-V$ decliners are
slow Lira law $B$ decliners\footnote{Note that this is different to
  the $B$--band decline rate directly after maximum, which is
  connected to different stretch or $\Delta m_{15}$ values}. The fact
that the Lira law $B-V$ decline rate is connected to the presence of
narrow absorption features suggests that this effect is either caused
by younger progenitor systems having slower Lira law $B$ decline
rates, which would be associated with star forming regions and larger
gas/dust column densities; or to the presence of circumstellar
material that causes the $B$ band evolution to become slower at late
times. We show the $B$ and $V$ band decline rates vs the $B-V$ decline
rates in Figure~\ref{fig:declines}.

\begin{figure*}[h!]
  \centering
  \hbox{
    \includegraphics[width=8cm,angle=0]{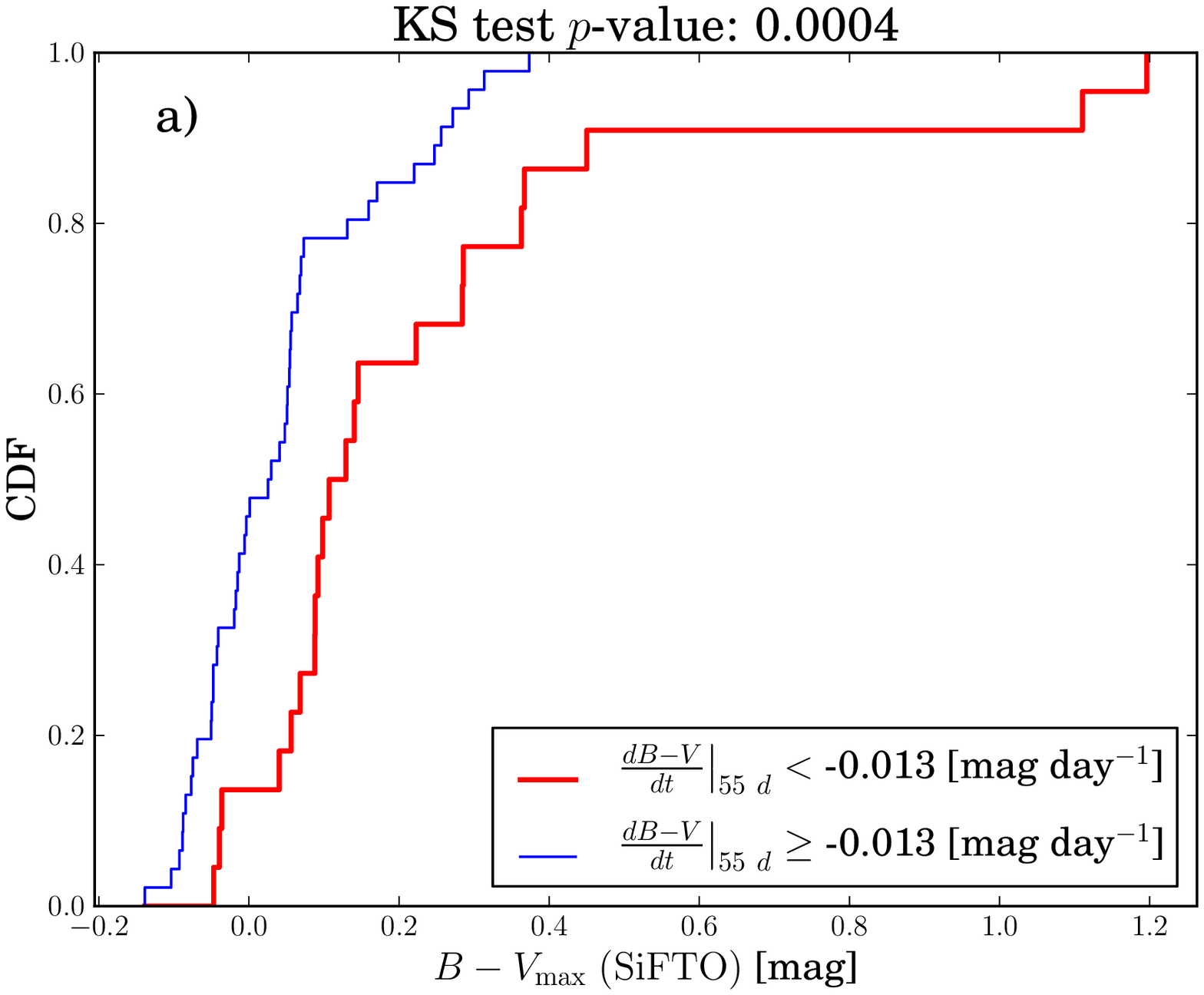}
    \includegraphics[width=8cm,angle=0]{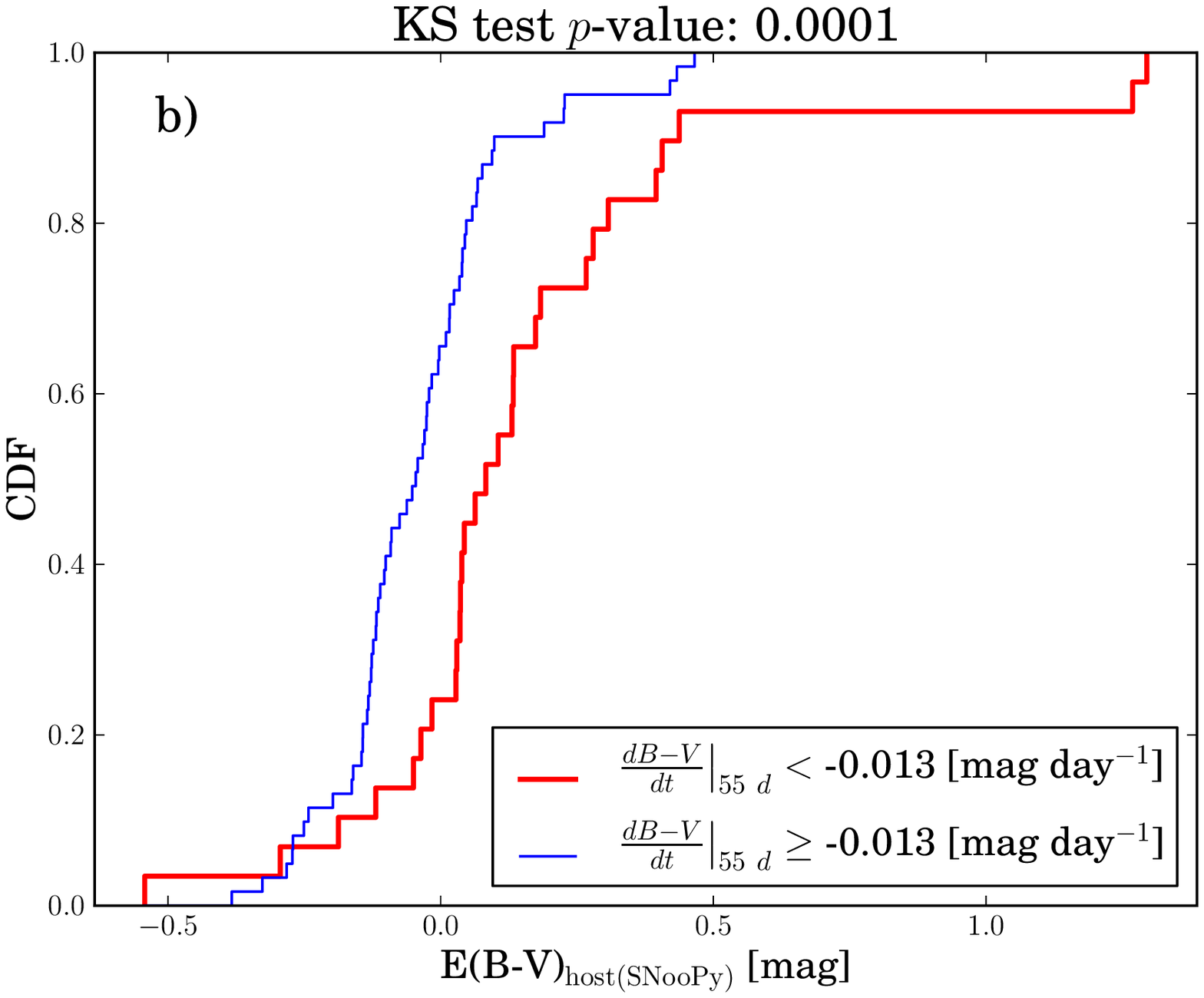} 
  }
  \caption{a) KS tests for the distributions of $B-V$ colors at
    maximum obtained from SiFTO dividing the sample by the Lira law
    $B-V$ decline rate and b) KS tests for the distributions of
    $E(B-V)$ extinction values obtained from SNooPy dividing the
    sample by the Lira law $B-V$ decline rate. Fast Lira law $B-V$
    decliners are redder at maximum, have stronger Na I D1 \& D2
    absorption, but have no strong preference for any particular
    environment (see Section~\ref{sec:env}), which is consistent with
    them having more CSM.}
  \label{fig:bBVcolormax}
\end{figure*}

If faster Lira law $B-V$ decliners are associated with more material
in the line of sight, we should also expect a difference in the
distributions of colors around maximum light of fast and slow Lira
decliners. In fact, the distributions of $B-V$ colors at maximum
obtained with SiFTO and $E(B-V)$ extinctions obtained from SNooPY are
significantly different if we divide the sample between fast and slow
Lira law $B-V$ decliners, with $p$--values of 0.0004 and 0.0001,
respectively. This is shown in Figure~\ref{fig:bBVcolormax}. Fast and
slow Lira law $B-V$ decliners also differ significantly in their
colors 55 days after maximum ($p$--value of 0.003).  It should be
noted that fast and slow Lira law $B-V$ decliners tend to have more
similar colors at late times during the Lira law, as redder SNe in
$B-V$ tend to decrease faster in $B-V$. These results are summarized
in Table~\ref{tab:results}.


\begin{table*}
  \centering
  \caption{KS test
  for the distribution of observed properties dividing the sample by
  Lira law $B-V$ decline rates. The null hypothesis is that fast and
  slow Lira law $B-V$ declining SNe Ia have the same distribution of
  blended Na I D1 \& D2 narrow absorption features, colors at maximum
  (SiFTO), extinction values (SNooPY) or colors 55 days after
  maximum. } 
  \begin{tabular}{ccccc}
    \hline
    \hline
    \vspace{-.3cm} \\
    KS test & EW(Na) & $B-V$ at max. & $E(B-V)$ & $B-V$ at 55 days \\
    \vspace{-.4cm} \\
    \hline
    $p$--value & 0.0006 & 0.0004 & 0.0001 & 0.003 \\
    \hline
  \end{tabular}
 \label{tab:results}
\end{table*}

\subsection{Environmental and SN intrinsic effects} \label{sec:env}

After finding that differences in the measured Lira law $B-V$ decline
rates are related to the presence of material in the line of sight, we
tested whether this could be explained by faster Lira law $B-V$
decliners occurring in younger stellar environments, i.e. whether the
material detected could be ISM. In order to do this we divide our
sample by quantities that should favor higher ISM column densities,
such as host galaxy morphological type or host galaxy inclinations
(removing E/S0s from the sample, where it is difficult to define an
inclination). If there was a strong preference for morphological type
among different Lira law $B-V$ decline rates it could mean that
younger systems decay faster during the Lira law, and that there is no
need to invoke CSM as an explanation. If there was a strong preference
for galaxy inclination, it would suggest that the Lira law $B-V$
decline rate is affected by ISM in the line of sight.

Stretch and $\Delta m_{15}$ are also thought to be associated with the
age of the progenitor systems \citep{1995AJ....109....1H,
  1996AJ....112.2398H, 1999AJ....117..707R, 2000AJ....120.1479H,
  2006ApJ...648..868S}, which we also confirmed in our sample, with
$p$--values of less than $0.00001$ for the null hypothesis that the
distributions of stretch and $\Delta m_{15}$ dividing the sample
between earlier and later type hosts arise from the same parent
population. Thus, stretch and $\Delta m_{15}$ play a similar role as
the host galaxy morphological type in the sense that younger systems
(larger stretch or smaller $\Delta m_{15}$) occur in regions where the
ISM column density should be larger. Therefore, if stretch or $\Delta
m_{15}$ were connected to the evolution of $B-V$ at late times it
could suggest that environment could drive the different Lira law
$B-V$ decline rates. On the other hand, stretch and $\Delta m_{15}$
are known to account for most of the diversity of SN Ia light curves
because they are likely related to the enclosed mass of $^{56}$Ni or
iron group elements (IGE) in the ejecta, which is the power source of
SN Ia light curves. Considering that at late times the physics of the
decay of $^{56}$Co and the absorption of positrons start to become
more relevant, it is not unreasonable to think that both stretch and
$\Delta m_{15}$ could be related to the Lira law $B-V$ decline rate.

Thus, we test whether the distributions of host morphological type,
host inclination, stretch and $\Delta m_{15}$ of fast and slow Lira
law $B-V$ decliners arise from the same parent population. In none of
the tests we obtained strongly significant differences (see
Table~\ref{tab:host}). Therefore, assuming that the Lira law $B-V$
decline rate is not strongly affected by the amount of ISM in the line
of sight, we interpret the results shown in Table~\ref{tab:results} as
evidence for CSM around those SNe Ia having the fastest Lira law $B-V$
decline rates (but see Section~\ref{sec:ISM}).

The presence of CSM could easily change the slope of the Lira law
$B-V$ decline rate, either by multiple scattering processes
\citep[i.e. CSM light echoes, see][]{2011ApJ...735...20A} or by the
progressive destruction of nearby dust if the material is sufficiently
close to the expanding ejecta, making the $B-V$ colors increasingly
bluer at late times and, as a consequence, the Lira law $B-V$ decline
rates faster. This was also noted by \citet{2010AJ....139..120F} in
systems with color excesses that were moderate to large.

\begin{table*}
  \centering
  \caption{KS test for the
  distribution of observables dividing the sample by Lira law $B-V$
  decline rates. The null hypothesis is that fast and slow declining
  Lira law SNe Ia have the same distribution of host morphological
  types, host inclinations, stretch or $\Delta m_{15}$. The null
  hypothesis cannot be discarded in any of these tests, suggesting
  that the equivalent width and color differences seen in
  Figure~\ref{fig:bBVew} are not driven by environmental factors.}
  \begin{tabular}{rcccc}
    \hline
    \hline
    \vspace{-.3cm} \\
    KS test & Host morph. type & Host. incl. & stretch & $\Delta m_{15}$ \\
    \vspace{-.4cm} \\
    \hline
    $p$--value & 0.10 & 0.19 & 0.18 & 0.07 \\
    \hline
  \end{tabular}
 \label{tab:host}
\end{table*}

If the $B-V$ colors at maximum are mainly driven by CSM reddening and
if either light echoes or late time CSM dust destruction occur, we
would expect that the reddening observed during maximum light should
decrease at later times, making the reddening to be first dominated by
CSM and later by ISM. In fact, we cannot rule out that the the
distributions of colors at maximum in the earlier and later type hosts
-- or lower and higher inclination hosts -- arise from the same parent
population ($p$--values of 0.16 and 0.39, respectively), as also found
in previous studies (see e.g. \citealt{2012arXiv1211.1386J}, c.f.
\citealt{2010MNRAS.406..782S}). This would be expected if the colors
of SN Ia at maximum were primarily driven by ISM.  However, we do rule
out that the distributions of colors 55 days after maximum in earlier
and later type hosts arise from the same parent population ($p$--value
of 0.003). This suggests that: 1) $B-V$ colors at maximum and 55 days
after maximum are driven by different physical effects, and 2) $B-V$
colors 55 days after maximum are more related to the properties of the
environment where SNe occur. This is consistent with the CSM
interpretation presented above when the Lira law decline rates are
considered. These results are summarized in Table~\ref{tab:BVmax55}.

\begin{table*} 
  \centering
  \caption{KS test for the
  distribution of $B-V$ colors at maximum ($B-V~_{\max}$) and 55 days
  after maximum ($B-V~_{\rm 55}$), dividing the sample by Lira law
  $B-V$ decline rates (columns 1 and 2) or by the median morphological
  type (columns 3 and 4). The null hypothesis is that fast and slow
  Lira law $B-V$ declining SNe Ia have the same distribution of colors
  at maximum light and 55 days after maximum light, or that earlier
  and later type hosts have the same distribution of colors at maximum
  and 55 days after maximum light. These results suggest that the
  reddening effect of CSM is more apparent at early times and that of
  ISM, at late times.}
  \begin{tabular}{ccccccccc}
    \hline
    \hline
    \vspace{-.3cm} \\
    KS test & & & $B-V~_{\max}$ & $B-V~_{55}$ & & & $B-V~_{\max}$ & $B-V~_{55}$ \\
    \hline
    Sample division & & & \multicolumn{2}{c}{Lira law $B-V$ decline rate} & & & \multicolumn{2}{c}{Morphological type} \\
    $p$--value & & & 0.0004 & 0.003 & & & 0.16 & 0.003 \\
    \hline
  \end{tabular}
 \label{tab:BVmax55}
\end{table*}

We should emphasize that we are not suggesting that ISM absorption is
irrelevant at maximum light, but only that it does not drive the
results shown in Table~\ref{tab:results} (although see
Section~\ref{sec:ISM}). In fact, we detect a highly significant
increase on the equivalent widths of Na I D1 \& D2 due to ISM related
properties by performing KS tests on the distribution of EWs of Na I
D1 \& D2 dividing a bigger sample by host galaxy type, host galaxy
inclination, stretch or $\Delta m_{15}$ instead of by Lira law $B-V$
decline rate, with $p$--values smaller than 0.00001 in all cases.

\subsection{$V-i$ colors and dust reddening laws ($R_{V}$)} \label{sec:rv}

\begin{figure*}[h!]
  \centering \hbox{ \includegraphics[width=8cm,angle=0]{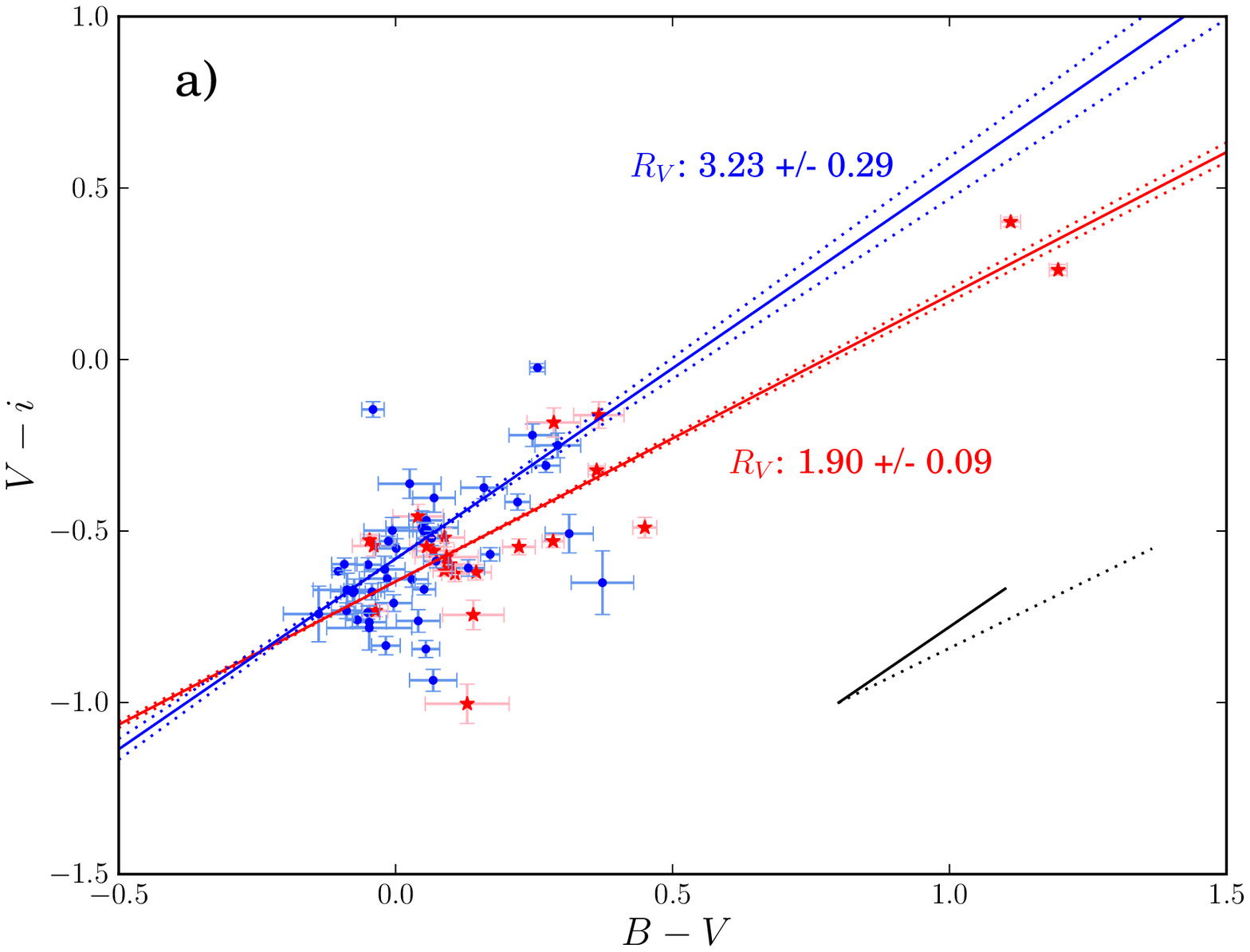}
    \includegraphics[width=8cm,angle=0]{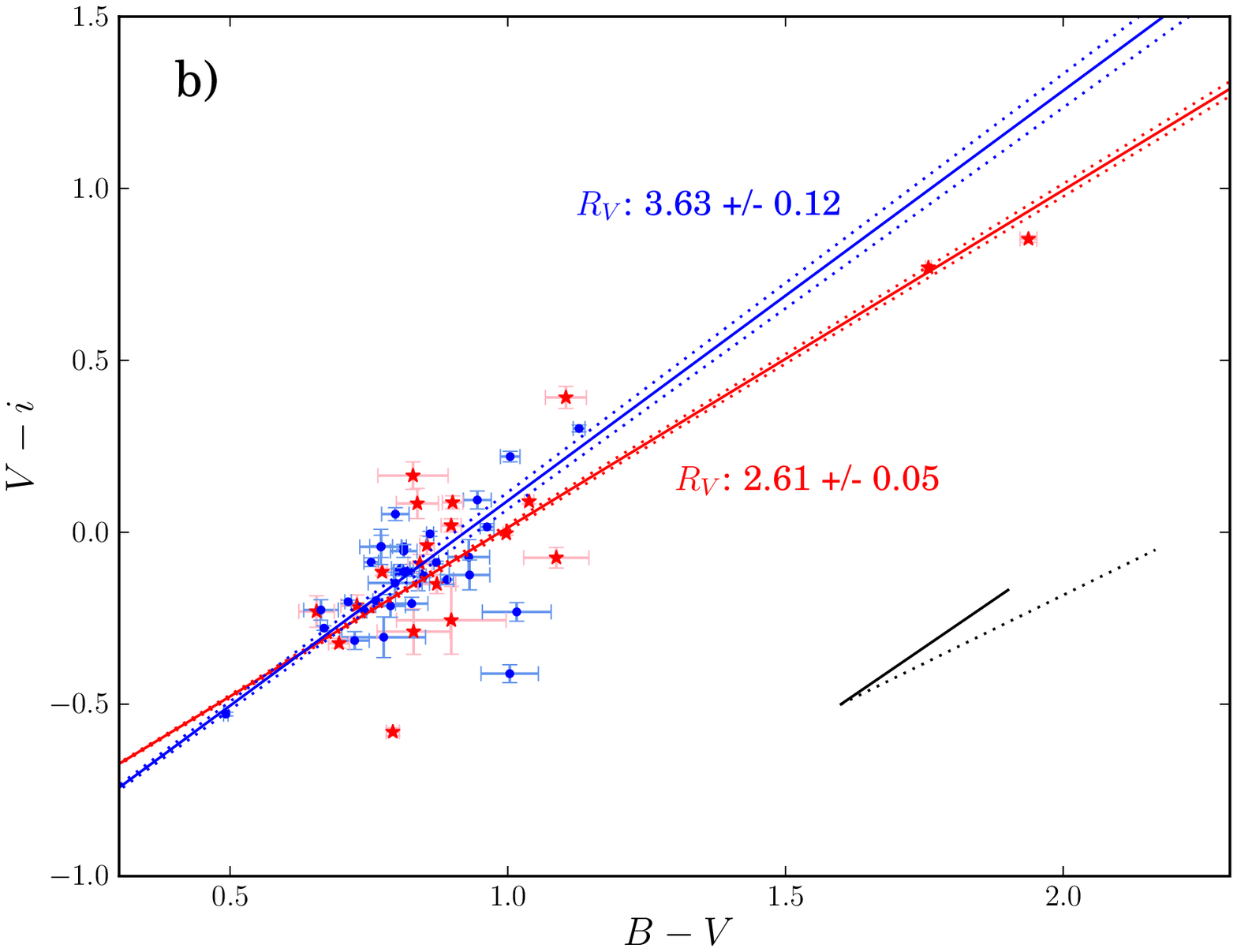} }
  \caption{a) $V-i$ and $B-V$ color at maximum light and b) 55 days
    after maximum light, dividing the sample between fast (red stars)
    and slow (blue dots) Lira law $B-V$ decliners. We fit a straight
    line to the $V-i$ vs $B-V$ colors at maximum light and 55 days
    after maximum for each sample. At maximum light and 55 days after
    maximum the extinction laws are significantly different, with
    significantly lower $R_{V}$ for fast Lira law $B-V$ decliners at
    all times, although becoming more similar to slow Lira law $B-V$
    decliners 55 days after maximum light. We show for reference
    extinction vectors with $A_{V} = 1$ for $R_{V} = 3.2$ (continuous)
    and $R_{V} = 1.7$ (dashed). If we remove the highest extinction
    SNe from the fast declining sample we obtain an even lower $R_{V}$
    at maximum.}
 \label{fig:Rvs}
\end{figure*}

A prediction from \citet{2008ApJ...686L.103G} is that those SNe Ia
with more CSM should have a steeper reddening law towards the blue.
Thus, we expect that fast Lira law $B-V$ decliners should have a lower
$R_{V}$ at maximum light if CSM was present in these systems, which
can be measured using $V-i$ colors. In Figure~\ref{fig:Rvs} we show
the $V-i$ vs $B-V$ colors at maximum and 55 days after maximum with
best--fitting $R_{\rm V}$ values assuming the reddening law of
\citet{1989ApJ...345..245C} and \citet{1994ApJ...422..158O}. To
estimate the $R_{V}$ best--fitting errors we have done a Montecarlo
analysis based on the measured $B$, $V$ and $i$ magnitudes and their
errors. We created 1000 realizations of these magnitudes at maximum
and 55 days after maximum to derive $B-V$ and $V-i$ colors, which by
definition are correlated, and then derive 1000 best--fitting $R_{\rm
  V}$ values, whose standard deviation was used as the error.

We find that fast Lira law decliners have a steeper reddening law
(lower $R_{V}$) than slow Lira law decliners, a result that becomes
more pronounced at maximum light, even after removing the highly
reddened SN 2003cg and SN 2006X. In fact, the $R_{\rm V}$ values that
best match the data for fast and slow Lira law decliners are $R_{\rm
  V} = 1.90 \pm 0.09$ and $R_{\rm V} = 3.23 \pm 0.29$, respectively,
which is consistent with what has been found for highly reddened and
normal SNe Ia \citep{2010AJ....139..120F, 2011ApJ...731..120M}. At 55
days after maximum we find that the reddening laws of fast and slow
Lira law decliners are significantly different, but more similar to
each other than at maximum light, with $R_{\rm V} = 2.61 \pm 0.05$ and
$R_{\rm V} = 3.63 \pm 0.12$ for fast and slow Lira law $B-V$
decliners, respectively. These results are consistent with a
contribution from light echoes and/or CSM dust destruction at late
times, which would make the $B-V$ colors bluer. However, we cannot
rule out an intrinsic relation between $V-i$ and $B-V$ colors at
maximum light and at late times that could mimic some of our results,
although this relation would need to be different for fast and slow
Lira law $B-V$ decliners.

We have also studied the $V-i$ colors at maximum and 55 days after
maximum of the SNe in the sample, in a similar fashion to $B-V$
colors. The probabilities that the distributions of $V-i$ colors at
maximum light or 55 days after maximum of fast and slow Lira law $B-V$
decliners arise from the same parent population is 0.13 and 0.07,
respectively. The lower significance obtained with these KS tests may
indicate that the effects we are measuring  are stronger at bluer
wavelengths.

\subsection{Robustness and systematic errors} \label{sec:stability}

\begin{figure*}[h!]
  \centering \includegraphics[width=14cm,angle=0]{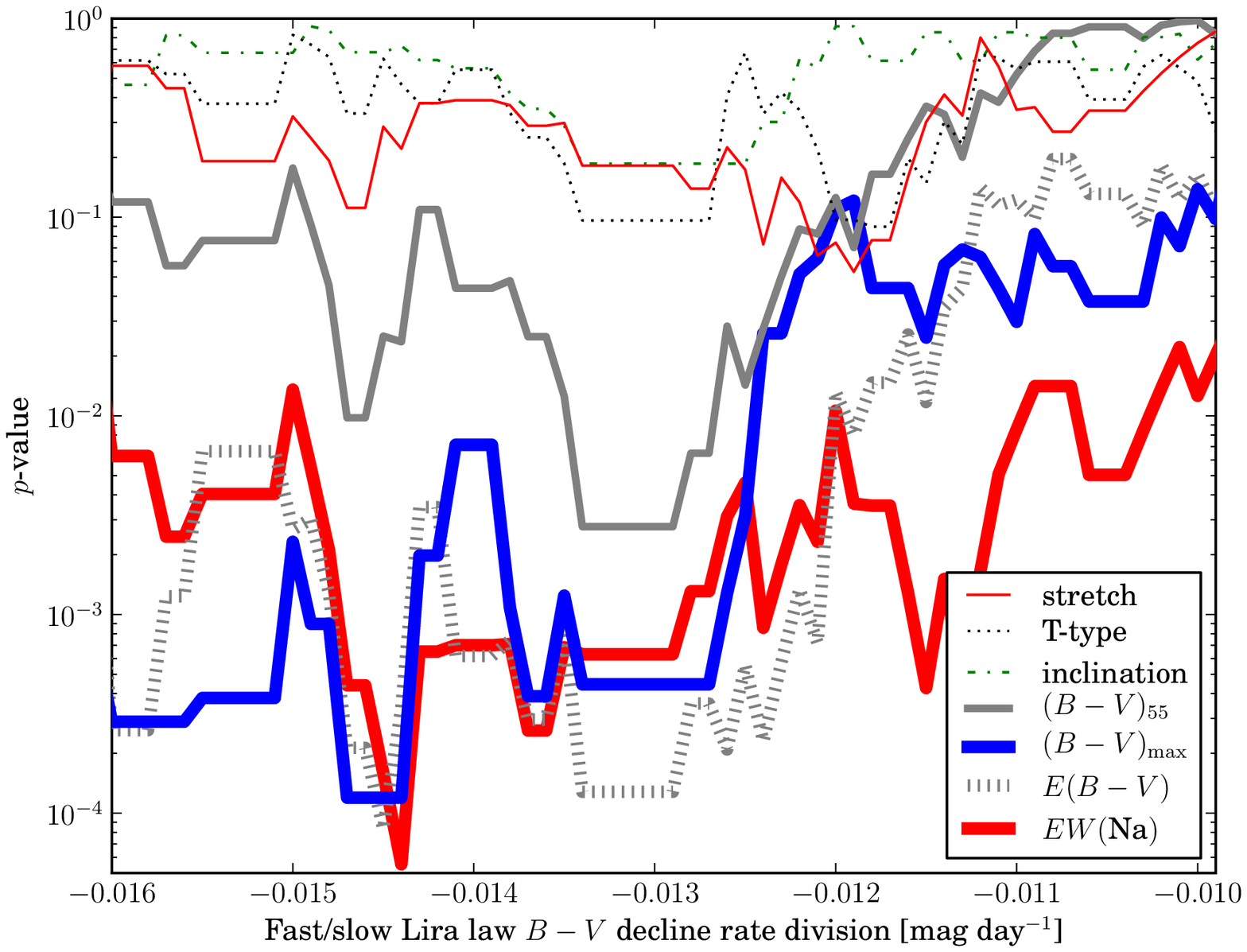}
  \caption{KS test $p$--values for the null hypothesis that the
    distribution of the variables indicated by the labels arise from
    the same parent population when dividing the sample by the Lira
    law $B-V$ decline rate shown in the $x$--axis. This plot shows
    that the distribution of colors at maximum and equivalent widths
    of blended Na I D1 \& D2 narrow absorption features have a robust
    correlation with the Lira law $B-V$ decline rate, and that the
    division of the sample between fast and slow Lira law decliners by
    a Lira law $B-V$ decline rate of -0.013 [mag day$^{-1}$] is
    appropriate.}
  \label{fig:pvalues}
\end{figure*}

In the previous analysis we set a seemingly arbitrary division of the
sample at a Lira law $B-V$ decline rate of $-0.013$ mag day$^{-1}$. In
order to define this limiting decline rate we performed a series of KS
tests for the distributions of equivalent widths of blended Na I D1 \&
D2 narrow absorption and $B-V$ colors. The optimal division was
defined by varying the dividing Lira law decline rate and finding a
small combined $p$--values of the KS--tests for the distributions of
equivalent widths and $B-V$ colors, which corresponded to a dividing
Lira law $B-V$ decline rate of about -0.013 mag day$^{-1}$. This is
shown in Figure~\ref{fig:pvalues}, where we also show the $p$--values
obtained for the KS test associated with some of the environmental
variables discussed before. This Figure shows that: 1) the equivalent
width of blended Na I D1 \& D2 narrow absorption features and the
colors at maximum and 55 days after maximum have a minimum $p$--value
at a similar dividing Lira law decline rate and 2) the environmental
variables considered in this analysis are never strongly associated
with different Lira law decline rates, independently of the division
of the sample used in the KS tests. Comparing the range of Lira law
decline rates considered with those found in Figure~\ref{fig:bBVew},
we see that the high significance is not very sensitive to the
dividing decline rate, making our conclusions more robust.

In order to check whether the results shown in the previous sections
could be an artifact due to different measurement errors, we performed
a KS--test on the distribution of the errors associated with the Lira
law $B-V$ decline rate among the fast and slow Lira law $B-V$
decliners, obtaining no significant differences between them
($p$--value = 0.34). In addition, in order to test whether our
conclusions could be affected by constraining the sample to SNe with
equivalent width errors lower than 0.6 \AA\ and to SNe with Lira law
$B-V$ decline rate errors lower than 0.006 mag day$^{-1}$, we relaxed
both constraints doubling the required errors and obtained consistent
results in all the relevant tests. Furthermore, to test whether fast
and slow Lira law decliners are distributed differently with distance,
which could affect some of our results through K--corrections, we
performed a KS--test on the distribution of the SN distances
separating the sample between fast and slow decliners, obtaining no
significant differences among the two groups ($p$--value = 0.49). To
test this possible bias differently, we restricted our sample to SNe
with distance modulus smaller than 35 mag and repeated all the
KS--test from previous sections, obtaining consistent results
again. Restricting the sample to only CfA3, CfA4 and CSP photometry we
also recover our main results, including different reddening laws
between fast and slow Lira law $B-V$ decliners.

We have also repeated our KS tests above after removing those SNe Ia
that are known to be extreme cases of either high column densities,
variable Na lines or peculiar ejecta properties: SN 2003cg, SN 2006X,
SN 2007le, SN 2000cx and SN 2005hk, and our main conclusions remain
the same. This indicates that the correlations found apply to a high
fraction of SNe Ia, and not only to those SNe Ia which have strong
CSM--like features like variable Na absorption.

Finally, in order to test whether the time interval used for measuring
the Lira law $B-V$ decline rate (35 to 80 days after maximum) affects
our conclusions, we have repeated the relevant tests using data from
different time intervals after maximum. Later time photometry could
have larger measurement errors and could therefore affect our best
fitting decline rates. Also, the choice of the start of the Lira law
regime is important since it could be too close to the observed peak
in color evolution. Thus, we measured decline rates with the following
time intervals after maximum: 30 to 90 days, 40 to 80 days, 30 to 70
days and 50 to 90 days. In all but the last interval, between 50 to 90
days, we recover our main results. This could be due to the effect of
larger errors or fewer data points at later times, or alternatively,
to a systematic flattening of the decline rate at later times in the
fast Lira law decliners.

\subsection{Nebular velocities} \label{sec:vneb}

\begin{figure*}[h!]
  \centering
  \includegraphics[width=12cm,angle=0]{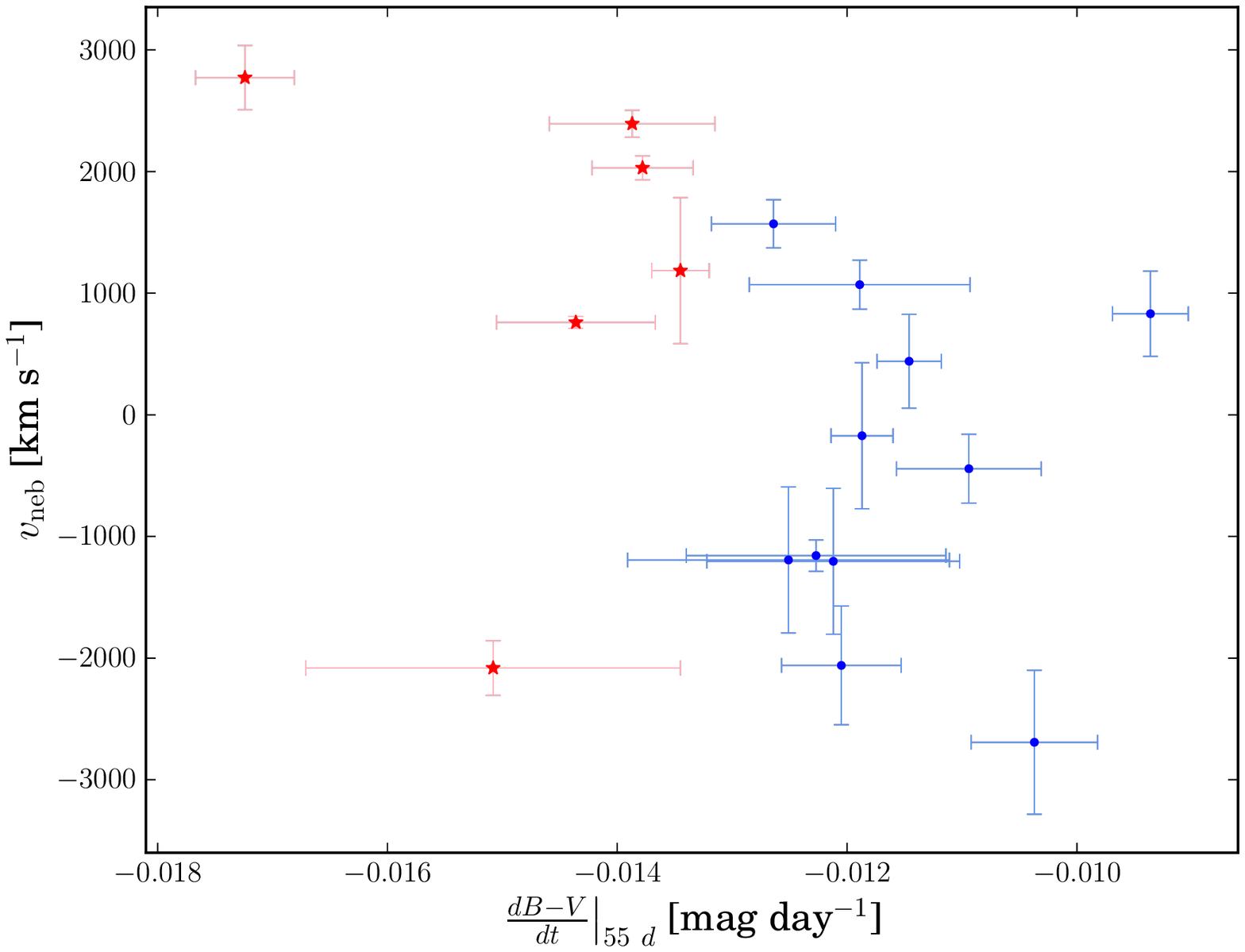}
  \caption{Nebular velocity shift of SNe Ia against their Lira law
    $B-V$ decline rate. Faster Lira law $B-V$ decliners appear to have
    positive nebular velocity shifts.  The fast declining SN with
    $v_{\rm neb} < 0$ is SN 1994D (see discussion in the text).}
 \label{fig:bBVvneb}
\end{figure*}

In FG12, a correlation between the presence of narrow absorption
features and positive nebular velocity shifts ($v_{\rm neb}$,
introduced by \citealt{2010ApJ...708.1703M}) was found, which was
interpreted as evidence for asymmetries in the CSM post explosion. In
order to test whether the correlations found by FG12 are consistent
with the results presented in this work, in Figure~\ref{fig:bBVvneb}
we show the nebular velocity shift of nearby SNe Ia against their Lira
law $B-V$ decline rate. For this we have taken a weighted average of
the nebular velocities from \citet{2012AJ....143..126B} and
\citet{2013MNRAS.430.1030S}. We could not reject the hypothesis that
the distributions of nebular velocities of fast and slow Lira law
$B-V$ decliners differ ($p$--value of 0.09). Nonetheless, the number
of SNe in the test is small and by changing the definition interval
for the Lira law $B-V$ decline rate from 35 to 80 days to 30 to 70
days we obtain a $p$--value of 0.003. This may suggest that SNe Ia
with the fastest Lira law $B-V$ decline rates tend to have positive
nebular velocity shifts. These are also the SNe found to have stronger
Na I D1 \& D2 absorption in FG12.

We note that in Figure~\ref{fig:bBVvneb} the fast declining SN with
$v_{\rm neb} < 0$ is SN 1994D, which is the only supernova in the
sample whose Lira law $B-V$ decline rate differs significantly from
the subtraction of its Lira law $B$ and $V$ decline rates, used in
Figure~\ref{fig:declines}.  If the latter is used, its decline rate
becomes much lower, -0.011 $\pm$ 0.002 mag day$^{-1}$ instead of
-0.015 $\pm$ 0.001, a difference which is explained by
non--simultaneous $B$ and $V$ photometry for some days, which cannot
be used to measure $B-V$ decline rates directly. If this different
value is used, we obtain a $p$--value of 0.02 for the probability that
the distributions of nebular velocities of fast and slow Lira law
$B-V$ decliners arise from the same population.

Given the correlation found by \citet{2010Natur.466...82M} between
nebular velocities and broad Si II $\lambda 6355 \AA$ absorption
velocity gradients, we investigated whether the Lira law $B-V$ decline
rates are related to the Si II absorption velocity at maximum. We
measured the Si II velocities at different epochs and obtained values
at maximum by interpolation, then combined these values with
measurements from other sources \citep{2005ApJ...623.1011B,
  2011ApJ...742...89F, 2012MNRAS.425.1789S, F+2013}, using weights
inversely proportional to the errors to obtain averages. We tested
whether Si II absorption velocities at maximum are related to Lira law
$B-V$ decline rates, colors at maximum (as found by
\citealt{2009ApJ...699L.139W} and \citealt{2011ApJ...742...89F}) or
equivalent widths of blended Na narrow absorption features, which
would be consistent with the excess blueshift found by
\citealt{2012ApJ...752..101F}. We found that there is a probability of
0.07 that the distributions of Lira law $B-V$ decline rates of SNe Ia
with Si II velocities faster or slower than 11,800 [km s$^{-1}$] arise
from the same parent population, a probability of 0.002 that the
distributions of colors at maximum of fast and slow Si II velocity SNe
Ia arise from the same parent population, a probability of 0.81 that
the distributions of equivalent widths of narrow Na absorption of fast
and slow Si II velocity SNe Ia arise from the same parent population
and a probability of 0.07 that the distributions of morphological
types of fast and slow Si II velocity SNe Ia arise from the same
parent population. This confirms the color differences found by
\citet{2009ApJ...699L.139W} and suggests a further connection with
Lira law $B-V$ decline rates. The non--detection of significant
differences in the distribution of equivalent widths of Na absorption
or morphological type with Si II velocity may indicate that some of
these color differences are intrinsic, or that our sample size is not
large enough, given the recent finding by \citep{2013arXiv1303.2601W}
of different radial distributions for fast and slow Si II velocity SNe
Ia.

\subsection{ISM as the origin of Lira law decline rate variations?} \label{sec:ISM}

Although the previous tests suggest that environmental factors do not
contribute to the Lira law $B-V$ decline rate, it is intriguing that
in Figure~\ref{fig:pvalues} the region where the $p$--values
associated with tests for a relation between Lira law decline rates
and the host galaxy morphological type or inclination are lowest
coincides with the region where $p$--values associated with equivalent
widths or colors are also the lowest. This led to us to consider
whether the ISM could really affect the Lira law $B-V$ decline rate in
some way.

Since the inclination of a galaxy should not be related to the
presence or absence of CSM in SNe Ia, and given that the ISM will be
generally too distant from the SN explosion to produce significant
light echoes or to be destroyed by the expanding ejecta between 35 and
80 days after maximum, a different mechanism to produce changes in the
Lira law $B-V$ decline rate would be needed. Note that we are not
referring to ISM light echoes \citep{2008ApJ...677.1060W}, which can
affect the light curves at even later times than those considered in
this work. In fact, the transversal expansion of the ejecta in an
inhomogeneous distribution of projected column densities of the host
galaxy ISM could affect the Lira law $B-V$ decline rate by changing
the average column density as the photospheric radius increases. If
this is the case, what needs to be shown is whether systems that have
stronger Na I D1 \& D2 absorption and redder colors at maximum are
also expected to have faster Lira law $B-V$ decline rates due to this
effect.

\begin{figure*}[h!]
  \centering \hbox{ \includegraphics[width=8cm,angle=0]{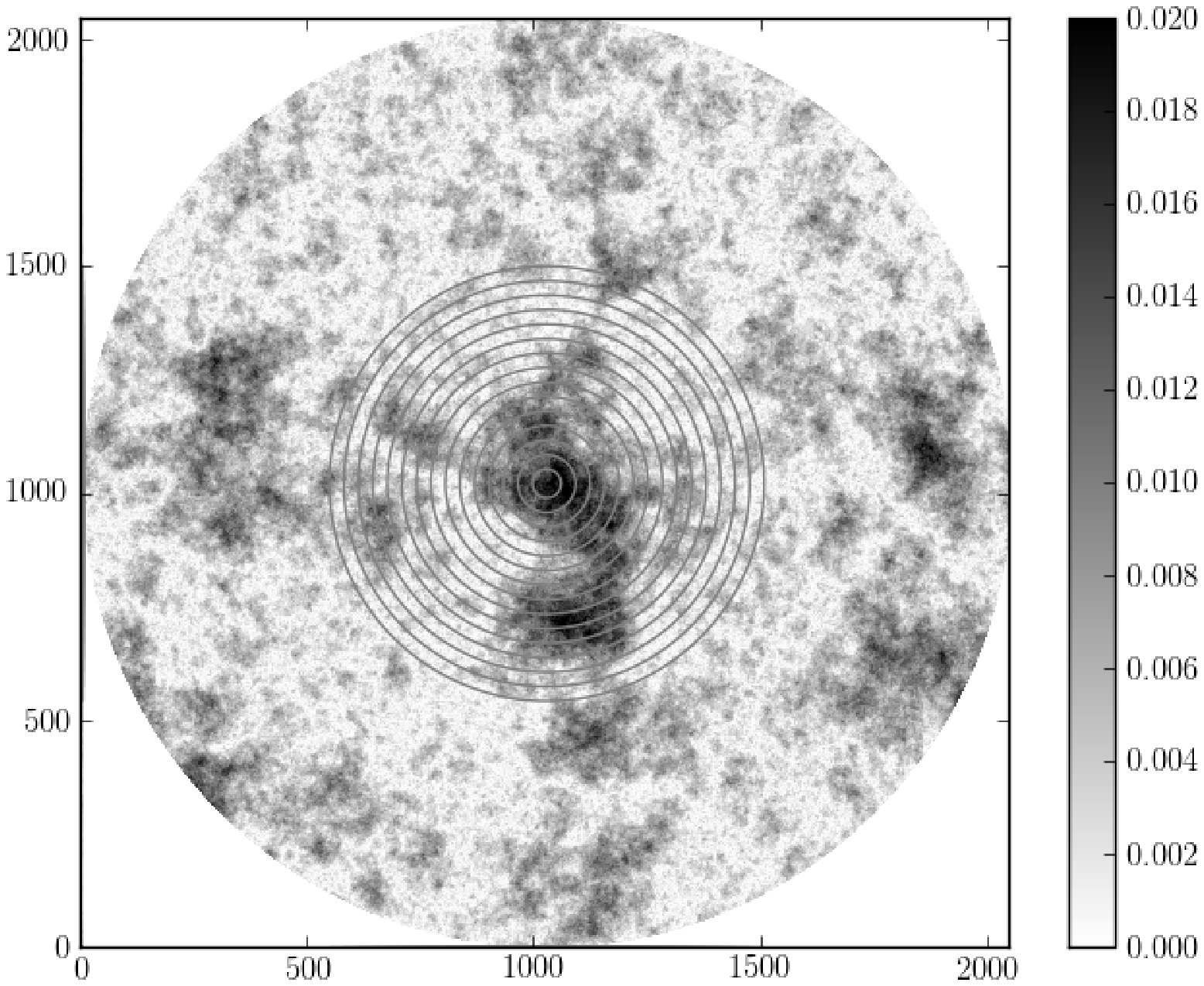}
    \includegraphics[width=8cm,angle=0]{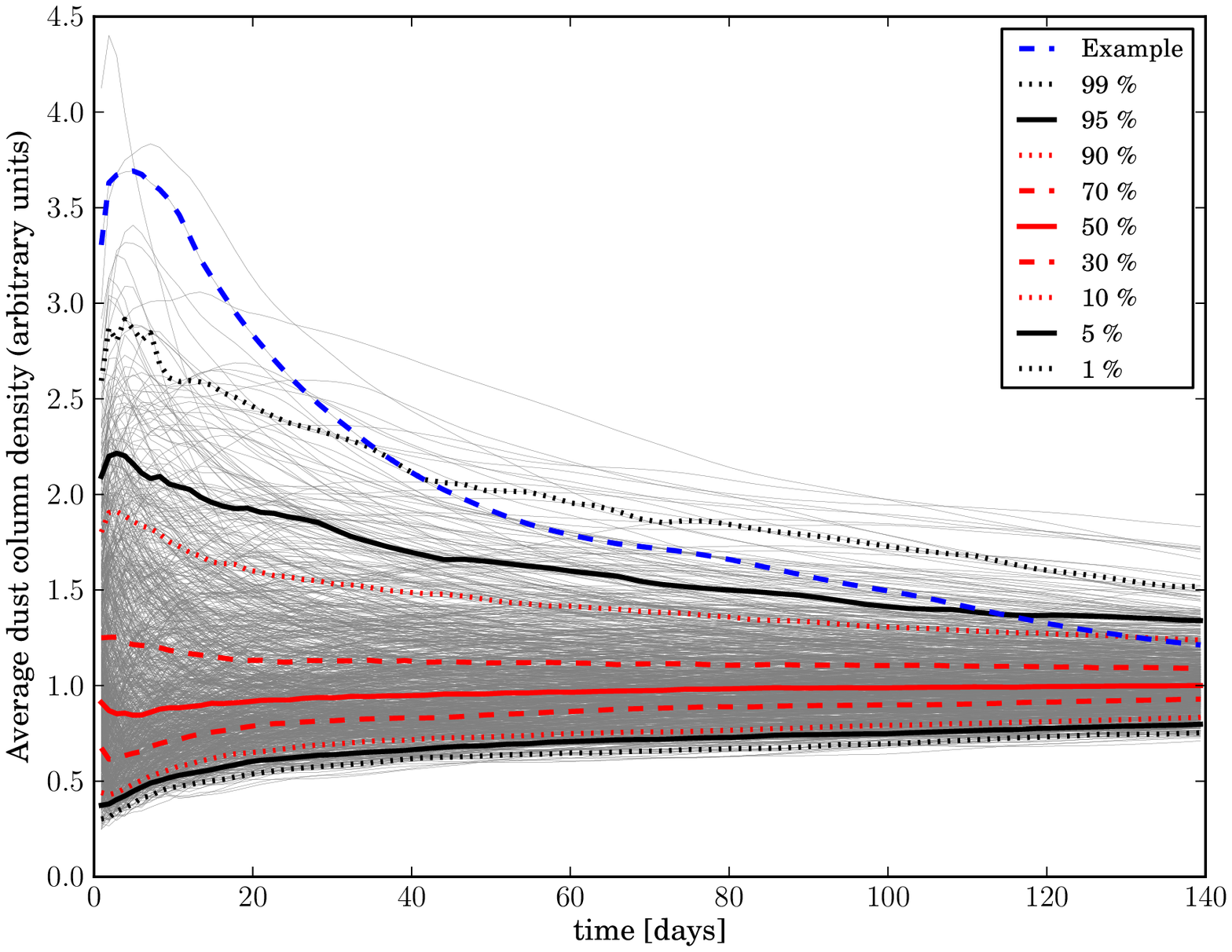} }
  \caption{\emph{Left:} example dust realization from the observer's
    perspective following the procedure of \citet{2010A&A...514A..78P}
    compared to the typical size of the SN Ia photosphere up to 100
    days after explosion. \emph{Right:} Evolution of the average
    column density within the photospheric disc of SNe Ia in 1000 ISM
    realizations. Most of the dust column density variations are found
    at earlier times. This means that the most reddened objects at
    earlier times will be less reddened at later times, implying a
    faster Lira law $B-V$ decline rate. The dashed--blue line
    corresponds to the example ISM realization shown, which is very
    atypical among the simulated ISM realizations.}
 \label{fig:ISM}
\end{figure*}

For this, we follow the approach of \citet{2010A&A...514A..78P} to
study the effect of the transversal expansion of the ejecta in a
non--homogeneous ISM. Assuming their approximation for the evolution
of the SN photospheric radius with time, and using random realizations
of the ISM column density assuming a fractal structure with power law
of -2.75 on spatial scales down to the size of the photosphere, we
compute the average column density within the photospheric radius as
the SN expands. In Figure~\ref{fig:ISM}, left, we show one realization
of the dust column density with different photospheric radii shown as
the SN expands. In this example, because the SN started behind a
relatively over--dense region, the average column density started at a
relatively large value and tended to decrease with time due to a
dilution effect as the photospheric radius increases. In
Figure~\ref{fig:ISM}, right, we show the expected evolution of the
average column density for one thousand ISM realizations, showing
different percentiles for reference. This figure shows that the
biggest dispersion occurs at early times, when chance alignments with
under or over--dense regions can dominate, but the dilution effect
occurring at later times will make the dispersion of the average
column densities decrease.

Thus, Figure~\ref{fig:ISM} shows that the inhomogeneous nature of the
ISM will naturally produce SNe Ia that have stronger equivalent widths
of narrow blended Na I D1 \& D2 absorption and are redder at maximum,
and that have fast--evolving average column densities, and therefore
faster Lira law $B-V$ decline rates. The higher the inclination, the
stronger the effect as the typical column density will tend to
increase. Moreover, at later times the colors will be dominated by the
overall column density, which could explain why SNe Ia in later
morphological types are significantly redder at later times. However,
this effect would not explain the differences in reddening laws
observed between fast and slow Lira law $B-V$ decliners.

Nonetheless, the column density is expected to be proportional to the
product of the color excess and $R_{\rm V}$, and a detailed analysis
of Figure~\ref{fig:ISM} implies that at maximum light the $B-V$ colors
of the SNe in the percentiles 50 and 95 should differ by about 0.25
mag\footnote{$-2.5 \log(\frac{n_{95}}{n_{50}}) / R_{\rm V} = -0.25$ assuming
  $R_{\rm V}= 3.2$, $n_{95} = 1.9$ units and $n_{50} = 0.9$ units.},
or by about 0.12 mag for the percentiles 30 and 70. At 55 days after
maximum, the $B-V$ colors of the SNe in the percentiles 50 and 95
should differ by about 0.15 mag, and by about 0.08 for the percentiles
30 and 70.  These magnitude differences are of the same order as those
seen in Figure~\ref{fig:BVsample}, but are not consistent with the
lack of correlation between the Lira law $V$--band and $B-V$ decline
rates observed in Figure~\ref{fig:declines}b. This suggests that the
inhomogeneous nature of the ISM can not explain all our findings, and
that a more detailed analysis will be needed to exclude this
possibility.

\section{DISCUSSION AND CONCLUSIONS}

In this work we have shown that there are significant differences in
the $B-V$ evolution of SNe Ia during the Lira law regime, between 35
and 80 days after maximum. These differences are driven by the
evolution of SNe Ia in the $B$--band: fast Lira law $B-V$ decliners
are slow Lira law $B$ decliners. With the sample defined in
Section~\ref{sec:sample} we have shown that fast Lira law $B-V$
decliners, defined as having Lira law $B-V$ decline rates smaller than
-0.013 mag day$^{-1}$, have stronger narrow absorption lines of
blended Na I D1 \& D2, are redder and have a a lower $R_{V}$
extinction law at maximum light and 55 days after maximum light. The
differences in colors were stronger at maximum light.

Additionally, we did not find a highly significant difference between
the distributions of host galaxy morphological type, host galaxy
inclination, stretch or $\Delta m_{15}$ dividing the sample by Lira
law $B-V$ decline rate.  This suggests that the Lira law $B-V$ decline
rates are not strongly dependent on the age of the SN Ia progenitor
systems, are not strongly affected by ISM, and are not mainly driven
by the the mass of nickel in the ejecta, the main parameter of SN Ia
light curves.

For the sample of 89 SNe Ia that had good photometric coverage up to
late times, we found that their colors at maximum were consistent
between earlier and later type galaxies. The colors 55 days after
maximum were significantly different between earlier and later type
galaxies, suggesting that the extinction at later times is more
clearly connected to ISM and that late time colors may be a better
proxy for environment. If CSM is dominant at early times, the
cancelling effect on the color excess due to CSM light echoes or to
dust destruction as the ejecta sweeps over the CSM could explain this
result.

A SN explosion surrounded by CSM should sublimate dust and ionize Na I
up to some maximum radii, therefore the CSM should have a
characteristic minimum radius, $r_{\rm CSM}$, which is thought to be
between 0.01 and 0.1 pc for Na I \citep{2007Sci...317..924P,
  2009ApJ...702.1157S}, and which is expected to be of the same order
of magnitude for dust \citep{2011ApJ...735...20A}. Photons leaving the
ejecta at maximum light could be seen as light echoes 55 days after
maximum if they are reflected after traveling for about 55 days, or a
typical distance of about 0.05 pc, which is of the order of this
minimum CSM radius. However, light echoes are not the only mechanism
that can produce this effect: depending on the outer ejecta speed,
which we assume to be one tenth of the speed of light, the ejecta
could have traveled up to about 0.001 pc at maximum light; therefore
it should not have time to interact with the CSM and sublimate CSM
dust. However, 55 days after maximum light the ejecta could be
approaching the inner edge of the CSM, since it will have traveled up
to about 0.006 pc, and it could start reducing the dust column
density.  This could explain why the differences in colors and
extinction laws found at maximum light between fast and slow Lira law
decliners were not as strong 55 days after maximum, but also why
environmental effects explain some of the differences in colors 55
days after maximum.

It is interesting to note that smaller dust sublimation or Na I D1 \&
D2 ionization CSM radii should produce more extinction and stronger
narrow absorption features, which we find in fast Lira law $B-V$
declining SNe Ia. For the same ejection velocity and mass loss rate
before a supernova explosion, the dust column density and extinction
should scale as $r_{CSM}^{-2}$. Smaller CSM dust inner radii should
also produce stronger CSM light echoes, since their intensity should
scale as $r_{CSM}^{-4}$, which would be reflected in faster $B-V$
evolution at late times because of the slower expected Lira law
$B$--band decline rates. In fact, \citet{2011ApJ...735...20A} showed
that smaller CSM radii should affect more significantly the $B-V$
evolution at later times, producing faster Lira law $B-V$ decline
rates.

The resulting fast declining sample is composed of 29 SNe Ia (33\%)
and the slow declining sample, by 60 SNe Ia (67\%). Interestingly, the
fraction of fast Lira decliners is similar to the fraction of 35\%
high velocity SNe Ia (HV SNe Ia, with $v(Si II)_{\max} \ge 11,800$ [km
  s$^{-1}$]) among the group of normal SNe Ia found by
\citet{2009ApJ...699L.139W}, or the excess of about 25\% of SNe Ia
with blueshifted absorption features detected by
\citet{2011Sci...333..856S}, which is also consistent with the excess
of blueshifted systems among HV SNe Ia found by
\citet{2012ApJ...752..101F}. This suggests that blueshifted narrow
absorption features, high velocity Si II features at maximum, high
equivalent widths of narrow absorption features, redder colors at
maximum, fast Lira law decline rates and lower $R_V$ reddening laws
may all be associated with the same population, which in our
interpretation is the one with significant CSM. Unfortunately, there
is no overlap between our sample and the recent sample of
CSM--interacting SNe Ia found by \citet{2013arXiv1304.0763S}. The
overlap between our sample and that from \citet{2011Sci...333..856S}
is very small (only 4 SNe) and we cannot test whether their
blueshifted sample corresponds to our fast declining sample. However,
the two overlapping SNe Ia in their blueshifted sample are also fast
Lira law $B-V$ decliners in our sample, SN 2006X and SN 2007le, and
they also happen to have variable Na I D1 \& D2 narrow absorption
features \citep{2007Sci...317..924P, 2009ApJ...702.1157S}, which is
expected in the CSM interpretation.

In the context of FG12, we have found a weak correlation between
nebular velocity shifts ($v_{\rm neb}$) and Lira law $B-V$ decline
rates, which is consistent with the interpretation of FG12, but which
needs to be confirmed with a bigger sample\footnote{We have confirmed
  the correlation between narrow absorption Na lines and nebular
  velocities using the public spectra from \cite{2013MNRAS.430.1030S}
  and the method described in FG12.}. This would mean that SNe Ia with
$v_{\rm neb} \ge 0$ have stronger equivalent widths of Na I D1 \& D2,
are redder at maximum light, have faster Lira law $B-V$ decline rates
and have lower $R_{V}$ reddening laws at maximum light because they
have more CSM in their line of sight. Assuming that about 33\% of the
SNe Ia have significant CSM in the line sight and that $v_{\rm neb}$
is a good proxy for explosion viewing angle, this would mean that
there is a comparable population of SNe Ia with negative nebular
velocities that would also have significant CSM, but which we do not
see in narrow absorption features or fast Lira law $B-V$ decline rates
because their CSM is on the opposite side of their ejecta. Here it
should be noted that dust scattering is primarily forward scattering
\citep{1941ApJ....93...70H}, which means that light echoes would be
best seen when the CSM is between the SNe and the observer ($v_{\rm
  neb} \ge 0$ in FG12). Although the fractions are uncertain at the
moment, this leaves room for additional progenitor scenarios that do
not fit in this CSM picture.

Since changes in the Lira law decline rate could be easily produced by
CSM \citep[][see their Figure 8]{2011ApJ...735...20A}, and given that
we did not find any significant environmental dependency of the Lira
law $B-V$ decline rates, we propose that the differences found between
fast and slow Lira law $B-V$ decliners are due to CSM, but we also
investigate the possibility that a similar result could be produced by
ISM. We showed that the transversal expansion of the SN photosphere
within an inhomogeneous distribution of column densities could
naturally lead to faster Lira law decliners having stronger absorption
at maximum as discussed in Section~\ref{sec:ISM}, but further
modeling is required to exclude this possibility.

This work is an addition to the growing evidence for CSM in SNe Ia,
but we believe that only the use of high resolution spectra
\citep{2011Sci...333..856S, 2012ApJ...752..101F} will help better
constrain the nature of the absorbing material in SNe Ia. The lack
of evidence for shocked winds, shocked companion stars, companion
stars in pre--supernova images or companion stars in SN Ia remnants
\citep[e.g.][]{2011Natur.480..348L, 2011Natur.480..344N,
  2012ApJ...744L..17B, 2012ApJ...753...22B, 2012Natur.481..164S,
  2012ApJ...747L..19E} may indicate that the progenitors of SNe Ia
arise from more than one progenitor scenario or from more complicated
scenarios that we have not yet imagined.

Finally, although it is still possible that the physics of positron
absorption could be responsible for the variations of the Lira law
detected, this effect should depend on the age of the progenitors in
order to produce stronger narrow absorption features due to larger ISM
column densities. The fact that we do not find a strong dependency
between Lira law decline rates and environment, stretch or $\Delta
m_{15}$ suggests that this is not the case. Additional studies of the
relation between very late $B-V$ decline rates (e.g. beyond 200 days)
and the Lira law $B-V$ decline rate could be used to provide a clean
separation between the effects of CSM and the physics of positron
absorption \citep[see e.g.][]{2006AJ....132.2024L}.

\section*{Acknowledgments}

\noindent

We thank an anonymous referee for useful suggestions that improved the
quality of the manuscript. We also thank M.~Phillips, G.~Pignata,
M.~Hamuy, J.~Anderson, F.~Bufano, P.~Zelaya, A.~Clochiatti, J.~Faherty
and P.~Rom\'an for many useful discussions. F.F. and S.G. acknowledge
support from FONDECYT through grant 3110042 and 3130680,
respectively. F.F and S.G. acknowledge support provided by the
Millennium Center for Supernova Science through grant P10-064-F
(funded by ``Programa Bicentenario de Ciencia y Tecnolog\'ia de
CONICYT'' and ``Programa Iniciativa Cient\'ifica Milenio de
MIDEPLAN''). G.~F. acknowledges support from the World Premier
International Research Center Initiative (WPI Initiative), MEXT,
Japan, and from Grant-in-Aid for Scientific Research for Young
Scientists (23740175). This research has made use of the CfA Supernova
Archive, which is funded in part by the National Science Foundation
through grant AST 0907903.

\end{document}